\documentclass[aps,prb,manuscript]{revtex4}
\usepackage{graphicx}
\usepackage{epsfig}
\usepackage{color}
\usepackage{dcolumn}
\usepackage{bm}
\usepackage{amsmath,amssymb}

\begin{document}
\draft
\title{Characterization of carrier transport properties in strained crystalline Si wall-like structures as a function of scaling into the quasi-quantum regime}

\author{C. S. Mayberry$^{1}$, Danhong Huang$^{1}$\footnote{Author to whom correspondence should be addressed. Electronic mail: danhong.huang\@us.af.mil},
G. Balakrishnan$^{2}$,\\ C. Kouhestani$^{1}$, N. Islam$^{3}$, S. R. J. Brueck$^{2,4}$, and A. K. Sharma$^{1,2}$}
\address{$^{1}$Air Force Research Laboratory, Space Vehicles Directorate, Kirtland Air Force Base, NM 87117, USA\\
$^{2}$Department of Electrical and Computer Engineering, University of New Mexico, Albuquerque NM 87106, USA\\
$^{3}$University of Missouri-Columbia, Columbia, MO 65211, USA\\
$^{4}$Department of Physics and Astronomy, University of New Mexico, Albuquerque NM 87106, USA}

\date{\today}

\begin{abstract}
We report the transport characteristics of both electrons and holes through narrow constricted
crystalline Si ``wall-like'' long-channels that were surrounded by a thermally grown SiO$_2$ layer. The
strained buffering depth inside the Si region (due to Si/SiO$_2$ interfacial lattice mismatch) is where scattering
is seen to enhance some modes of the carrier-lattice interaction, while suppressing others, thereby changing
the relative value of the effective masses of both electrons and holes, as compared to bulk Si. Importantly, as a
result of the existence of fixed oxide charges in the thermally grown SiO$_2$ layer and the Si/SiO$_2$
interface, the effective Si cross-sectional wall widths were considerably narrower than the actual
physical widths, due the formation of depletion regions from both sides. The physical height of the
crystalline-Si structures was $1500$\,nm, and the widths were incrementally scaled down from $200$\,nm to $20$\,nm.
These nanostructures were configured into a metal-semiconductor-metal device configuration that was
isolated from the substrate region. Dark currents, dc-photo-response, and carrier ``time-of-flight'' response
measurements using a mode-locked femtosecond laser, were used in the study. In the narrowest wall devices,
a considerable increase in conductivity was observed as a result of higher carrier mobilities due to lateral
constriction and strain. The strain effects, which include the reversal splitting of light- and heavy- hole bands
as well as the decrease of conduction-band effective mass by reduced Si bandgap energy, are formulated in
our microscopic model for explaining the experimentally observed enhancements in both conduction- and
valence-band mobilities with reduced Si wall thickness. The role of the biaxial strain buffing depth is
elucidated and the quasi-quantum effect for the saturation hole mobility at small wall thickness is also found
and explained. Specifically, the enhancements of the valence-band and conduction-band mobilities are
found to be associated with different aspects of theoretical model.
\end{abstract}
\pacs{PACS:}
\maketitle

\section{Introduction}
\label{sec1}

For over 40 years the microelectronics market place has driven the very large scale integration
(VLSI) industry to make continuous improvements in computational power, bandwidth and
speed.\,\cite{add1} These continued enhancements in performance have come in the form of
``cramming'' more components onto integrated circuits, as was predicted in 1965 by Gordon E. More.\,\cite{add2}
\medskip

The push to increase the speed and density of the transistors on a chip has come in the form of
shrinking the transistor size, in particular the channel length.\,\cite{add3,add4} However, reductions in channel
length have come with challenges {\em i.e.}, short channel effects.\,\cite{add5} Short channel effects lead to higher
leakage currents, poor signal-to-noise ratios and instability during operation, such as loss of
channel's gate control.
\medskip

In order to improve the transistor's gate control and switching speed, the contemporary metal–oxide–semiconductor (CMOS) industry has
looked for alternative solutions to the traditional planar transistor designs and substrates.\,\cite{add6}
\medskip

Over the past several years, the CMOS industry has narrowed their focus into multi-gate field
effect transistor designs\,\cite{add7} for improving the gate control, and strained substrates\,\cite{add8,add9} to enhance
carrier carrier mobilities and ultimately the switching speed and drive currents.
\medskip

One particular multigate transistor design that has gained considerable interest among the
industry, as a replacement for the planar design, is the FinFET\,\cite{add7}. The FinFET has a tri-gate
architecture and reductions of short-channel effects have been observed in these devices.\,\cite{add10,add11} This
design provides gate control, not only from the top of the channel, but also from the channel sides
as well. This in itself improves the overall (on/off) gate control process, however, the drawback is
that these devices require higher operating voltages to achieve faster switching speeds.\,\cite{add12} S. W.
Bedell {\em et. al.\/},\,\cite{r7} has reported that in the present FinFET technologies, the carrier mobilities are not
seen to be enhanced, since the active region of these devices would require two opposite kinds of
strain (i.e. tensile and compressive) on the same substrate, which would be possible by converting
the tensile strain of silicon-on-insulator substrates to compressive strain in localized regions via a
combination of selective SiGe($>40\%$) growth.
Such FinFET devices have not yet been experimentally demonstrated. 
However, similar embedded silicon/germanium layered structures that would provide process induced stressors 
in the source and drain regions have been theoretically modeled. The results of these simulations show
only a modest performance increase, approximately
one-half enhancement in mobility, as compared to similar size planar FETs.\,\cite{sun} Process induced
strain in a FinFETs would be most effective if it was directly under the gate region, as its
stressor's effectiveness diminishes with depth. Incorporation of wafer level strain using SiGe-on-insulator
(SGOI) and Strained Silicon-On-Insulator (sSOI) in small pitched circuits may be
possible by converting the tensile strain of sSOI to compressive strain by selective growth of
silicon-germanium. However, these type of configurations would certainly add complexities
in a high volume manufacturing environment, which could negatively affect yield.
\medskip

In this paper we report an comprehensive experimental and theoretical study on the nature of
carrier transport, of both electrons and holes, through narrow constricted crystalline Si ``wall-like''
long-channels that were surrounded by a thermally grown SiO$_2$ layer. The carrier transport
characteristics are evaluated as a function of dimensional scaling of the Si wall widths from
$200$\,nm to $20$\,nm. The Si wall-widths were reduced by the process of thermal oxidation, where
stress naturally accumulates in the channel. Basically, this structure configuration allows us to
investigate the effects of strained regions that are ``closing-in'' from both sides. Additionally, as the
wall-widths approach the quasi quantum regime, the carriers start to become confined and
therefore react to the narrow paths, and possibly behave more like waves then particles,\,\cite{add15} thus
altering the macroscopic nature of resistance, capacitance and inductance to a more exotic
microscopic one.\,\cite{add16} However, this transition into the quantum mechanical regime does not
come about abruptly. Rather, there is a transition region in which the bulk properties begin to
slowly weaken while the quantum effects begin to strengthen.
\medskip

The effects of quantum confinement on carrier transport properties, however, have been
primarily investigated in ternary and quarternary material heterostructures and supperlattices, in
which scattering is seen to enhance some modes of the electron-lattice interactions while
suppressing others, thereby changing the relative value of the carrier's effective masses of
electrons and holes, as compared to bulk semiconductors.\,\cite{r4} To date such studies in Si have been
very limited. We believe that these wall structures are a useful starting point for a broader study,
as these can be configured into novel high density 3-D VLSI devices, where thermal effects, such
as heat buildup, can also be efficiently managed.\,\cite{add18}
\medskip

We organize the rest of the paper as follows: In Sec.\,\ref{sec2} the process for fabricating the crystalline Si
wall structures is described, and the two electrode, metal-semiconductor-metal device structure
fabrication is also discussed. In Sec.\,\ref{sec3}, the experimental dc measurements as a function of wall
width thickness are presented, including dark currents and photocurrents.
\medskip

The electron and hole transient time responses and analysis are done in Sec.\,\ref{sec3}. Section\ \ref{sec4}
provides a detailed model to explain the role of strain effects and how they impact both the
electron and hole mobilities, while we summarize in Sec.\,\ref{sec5}.

\section{Fabrication}
\label{sec2}

\subsection{Rationale for substrate material}
\label{sec2.1}

Silicon-on-insulator (SOI) wafers with a top active layer of $<100>$ crystal orientation were
used to fabricate the wall-like structured devices for this study. The initial SOI structure had a $1500$\,
nm active layer on top of a $3000$\,nm buried oxide. The SOI configuration allowed complete
electrical isolation of the Si wall-like structures from the underlying substrate. All five samples
in this study had identical $p$-type active layer with a lightly doped concentration of $~10^{14}$\,cm$^{-3}$
boron atoms. Intrinsic SOI wafers would have been an ideal choice for the experiment;
however, due to the commercial unavailability of $100\%$ intrinsic material, the above choice of
dopant type and concentration was adequate enough to minimize the effects of impurity
scattering. Boron tends to segregate away from the Si interface and into the thermally grown
oxide\,\cite{add19}, thus reducing the impurity concentration near the Si interface with SiO$_2$. Thus the
segregation coefficient, which is defined as the ratio of the dopant concentrations at the
interface, is less than one in our case. The thermal oxidation process leads to the formation of an
oxide trapped charge ($Q_{\rm ot}$), which contribute to the formation of a depletion region near the Si/
SiO$_2$ interface.\,\cite{r9,r10} Now, if we combine this oxide trapped charge with the fixed charge ($Q_{\rm f}$)
which naturally results from the excess Si atoms not reacted with the oxygen, and the interface
trapped charge ($Q_{\rm it}$) which results from the mismatch between the number of atomic bonds in the
Si crystal surface and the number of available bonds in the SiO$_2$ layer, these all sum to ($Q_{\rm ot}+Q_{\rm f}+Q_{\rm it}$).
Combined these form an depletion region in Si that extends several nanometers away from
the SiO$_2$ interface.\,\cite{r10,r11} Thus the effective Si cross-sectional wall widths were considerably narrower
than the actual physical widths, due this formation of depletion regions from both sides.

\subsection{Crystalline silicon wall nanostructure fabrication}
\label{sec2.2}

\begin{figure}[p]
\centering
\includegraphics[width=0.3\textwidth]{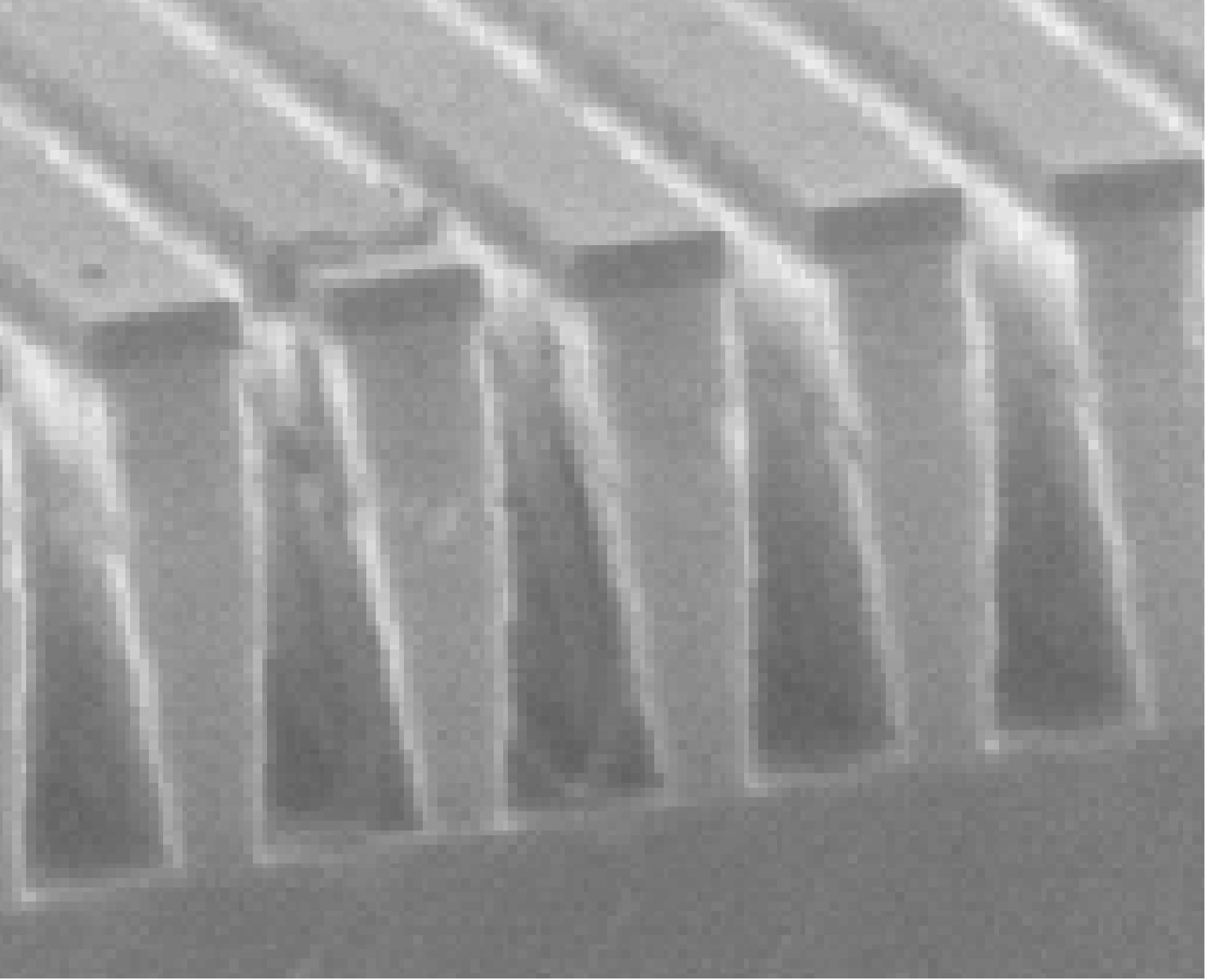}
\caption{\label{f1}
Scanning electron microscope (SEM) cross-sectional image of an array of wall precursor structures with a remaining layer of patterned photo-resist after reactive ion etch process step.}
\end{figure}

\begin{figure}[p]
\centering
\includegraphics[width=0.3\textwidth]{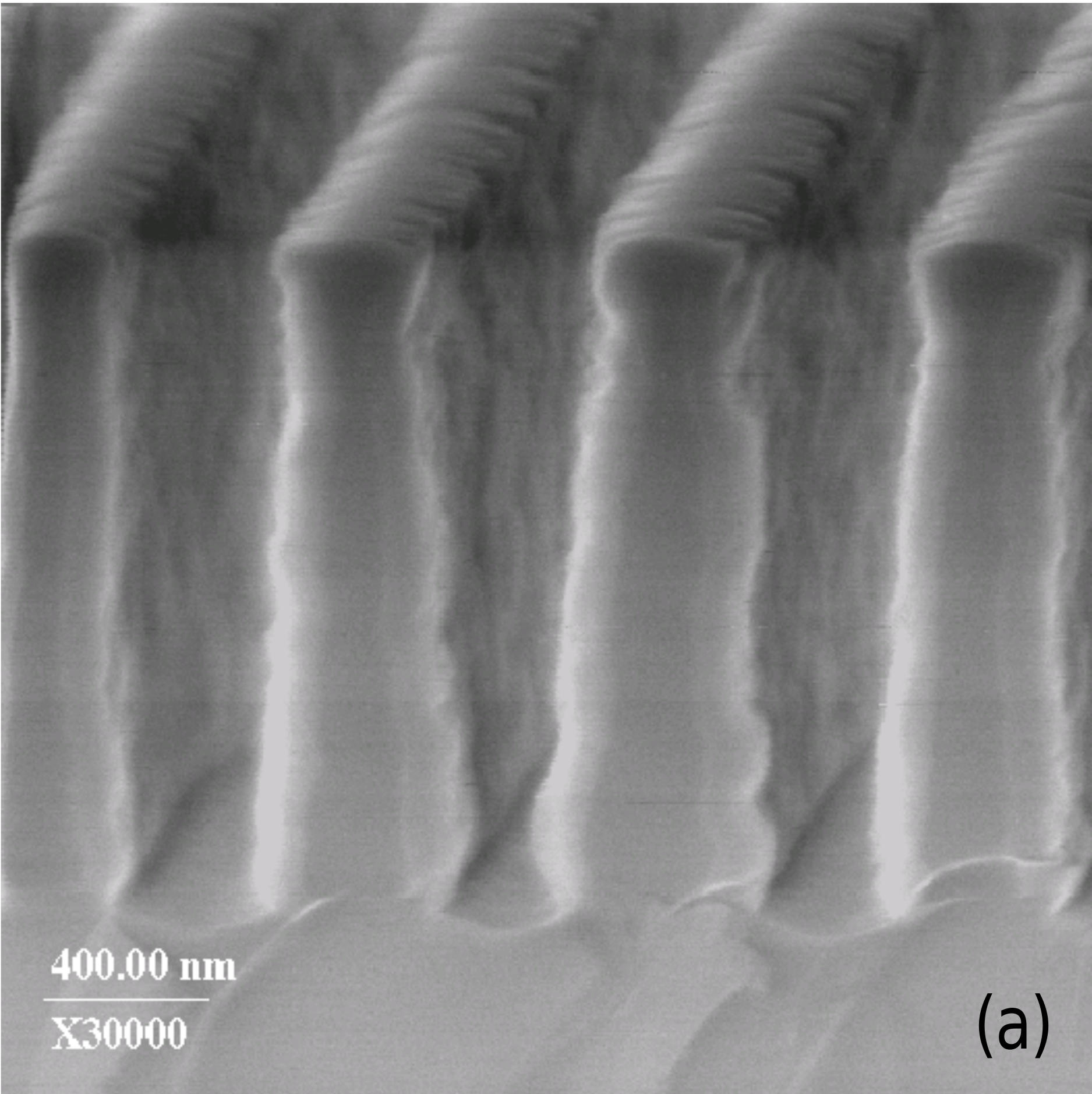}
\includegraphics[width=0.3\textwidth]{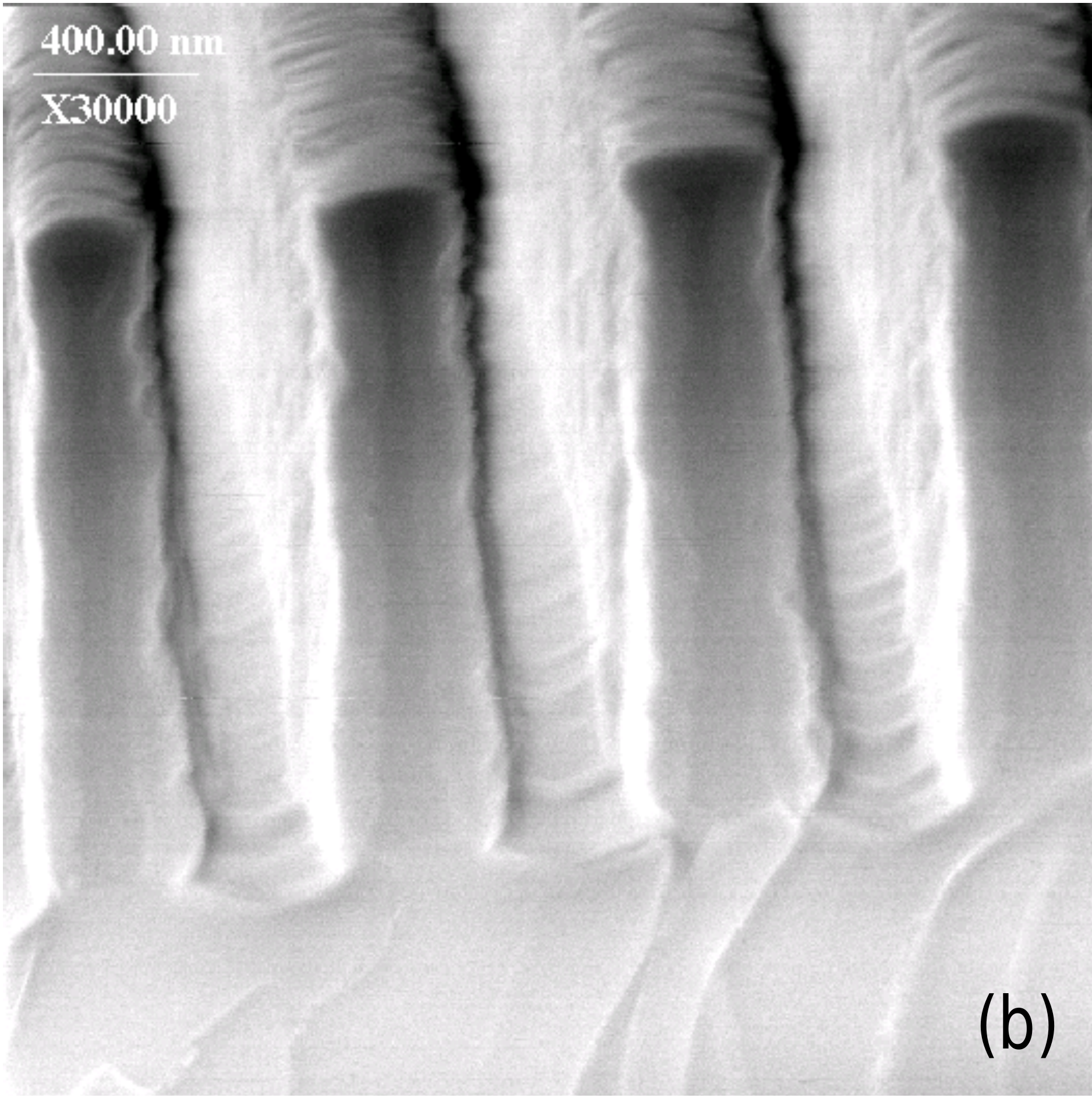}
\includegraphics[width=0.3\textwidth]{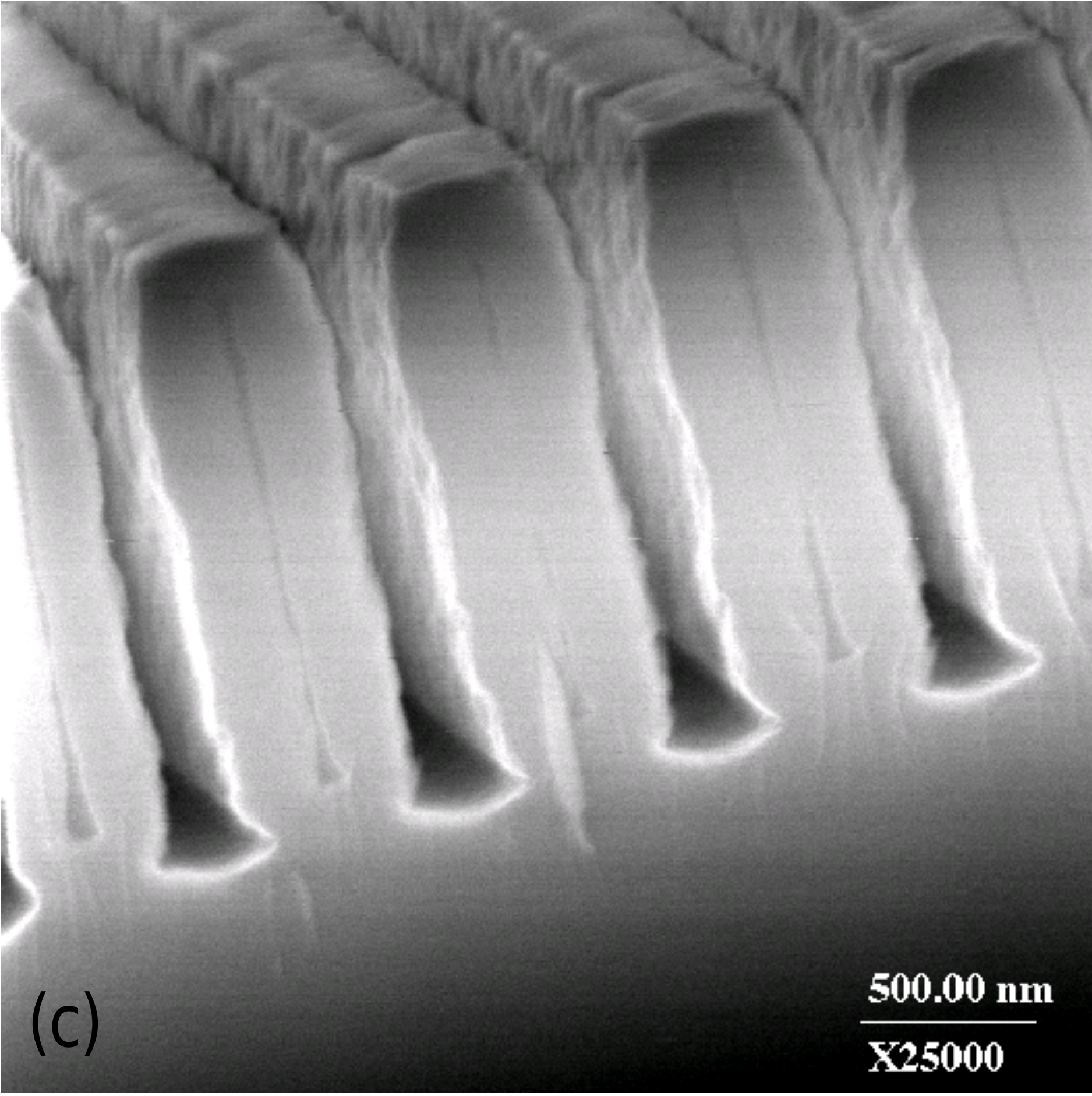}
\caption{\label{f2}
SEM cross-sectional images of an array of wall structures after thermal oxidation: (a) $~200$\,nm wall structures; (b) $~95$\,nm wall structures;
(c) $~40$\,nm wall structures.}
\end{figure}

In order to fabricate the wall structures, photoresist nano-scale pattering was required.
The precursors to the wall structures were patterned using interferometric lithography (IL)\,\cite{r15}
and reactive-ion-etching (RIE)\,\cite{add24,add25}. IL is a well-developed technique for inexpensive nano-
patterning process.\,\cite{add26} IL, in its simplest form, is interference between two coherent waves
resulting in a 1-D periodic pattern defined by $\lambda/2\sin\theta$ where $\lambda$ is the optical wavelength
and $2\theta$ is the angle between the interfering beams. A typical IL configuration consists of a
collimated laser beam incident on a Fresnel mirror (FM) arrangement\,\cite{add27} mounted on a
rotation stage for period variation. There is no $z$-dependence to an IL exposure pattern, which
is limited only by the laser coherence length and beam overlaps\,\cite{add28}. The 1-D nanoscale patterns
were first formed in the photoresist followed by pattern transfer onto the underlying substrate
using RIE in a parallel plate reactor using SF$_6$ plasma chemistry. Figure\ \ref{f1} shows a scanning
electron microscope (SEM) cross-sectional image of an array of nano-wall structures with a
remaining layer of patterned photoresist after RIE has been performed. Note at this stage
these structures are merely the precursors to the thin Si wall structures that are then
reconfigured into metal-semiconductor-metal (MSM) devices. After the photoresist was re-
moved, the wall structures are thermally oxidized. The oxidation process accomplished two
things. First, it consumes the Si, thus thins the wall width. Secondly, the thermally grown
oxide preserves a low defect, clean Si/SiO$_2$ interface, and at the same time passivates the surfaces of
the nanostructures.\,\cite{r15,r16} The Si/SiO$_2$ interface has low defects and it is important to note that
strain is present at the interface and it reduces with distance from the interface. This reduction
in strain as a f unction of depth has been seen experimentally in Si/SiO$_2$ interfaces using a
scanning transmission electron microscope using Z-contrast imaging which produces strain
contrast imaging\,\cite{r18}. Using this technique, the $1/e$ decay length was measured at approximately
$1$\,nm. The modeling of the thermal oxidation parameters needed for the desired thicknesses
was complicated due to the f act that in a three dimensional wall structure there are several
crystal lattice orientations that have different thermal oxidation rates. As a first order
approximation, we used average values of oxidation rates between the various lattice
orientations {\em i.e.\/} oxygen
ow rate, pressure, temperature and time. These parameters were
then fine-tuned empirically during the actual thermal oxidation runs. Figure\ \ref{f2}(a)-\ref{f2}(c) show
SEM images of the cross-sectional views of the wall structures after the respective thermal
oxidations. As can be seen from the SEM images, due to the high aspect ratio of these
structures the oxidation rate was not fully uniform throughout the height of the walls. The
rate was faster at the top part of the walls and slower at the bottom part due to higher
availability of oxygen atoms in the upper regions. The resulting wall-like Si structures
surrounded by the thermally grown oxide are then configured into the active region of the MSM
devices as described in the next section.

\subsection{MSM device fabrication}
\label{sec2.3}

\begin{figure}[p]
\centering
\includegraphics[width=0.3\textwidth]{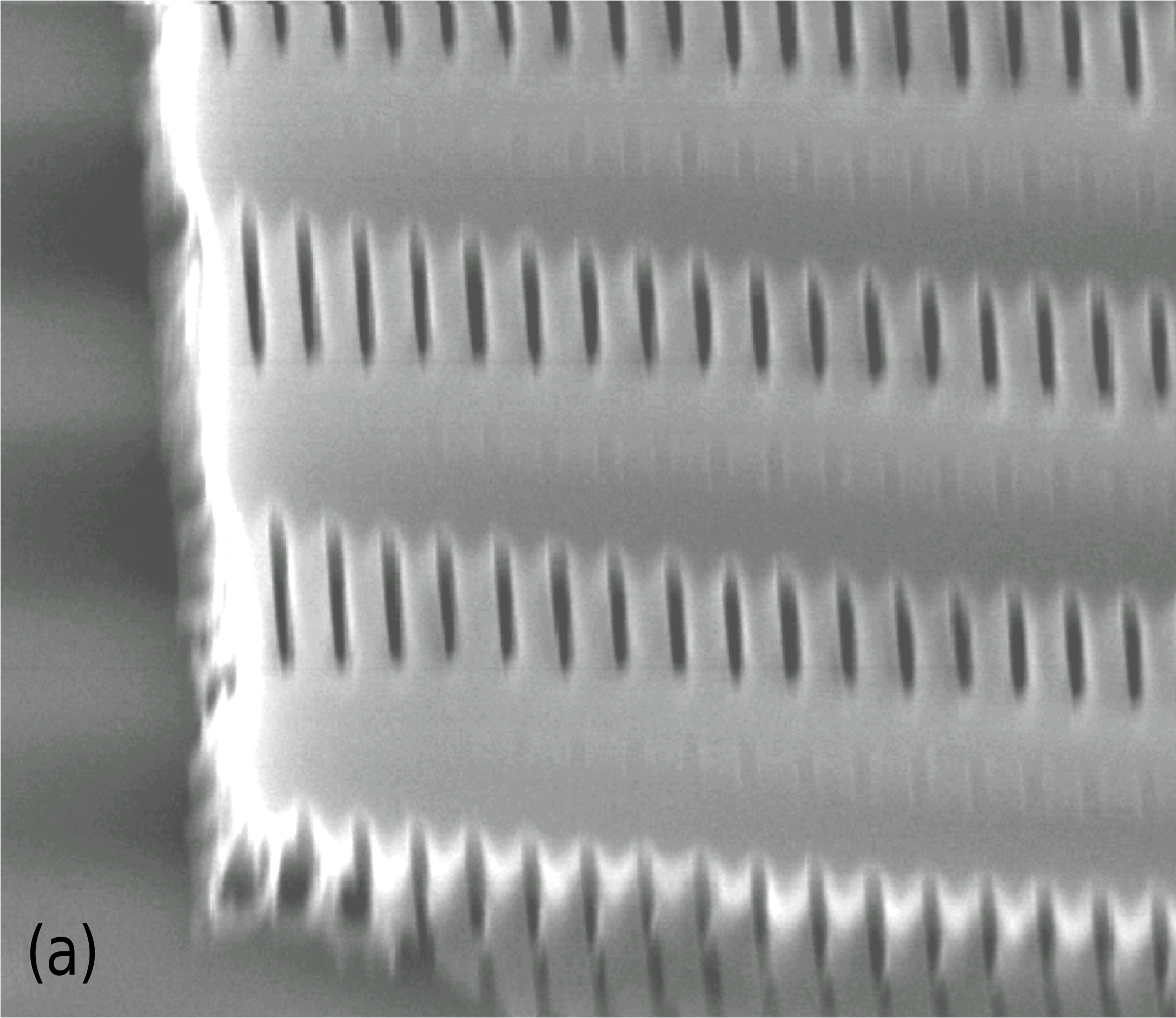}
\includegraphics[width=0.3\textwidth]{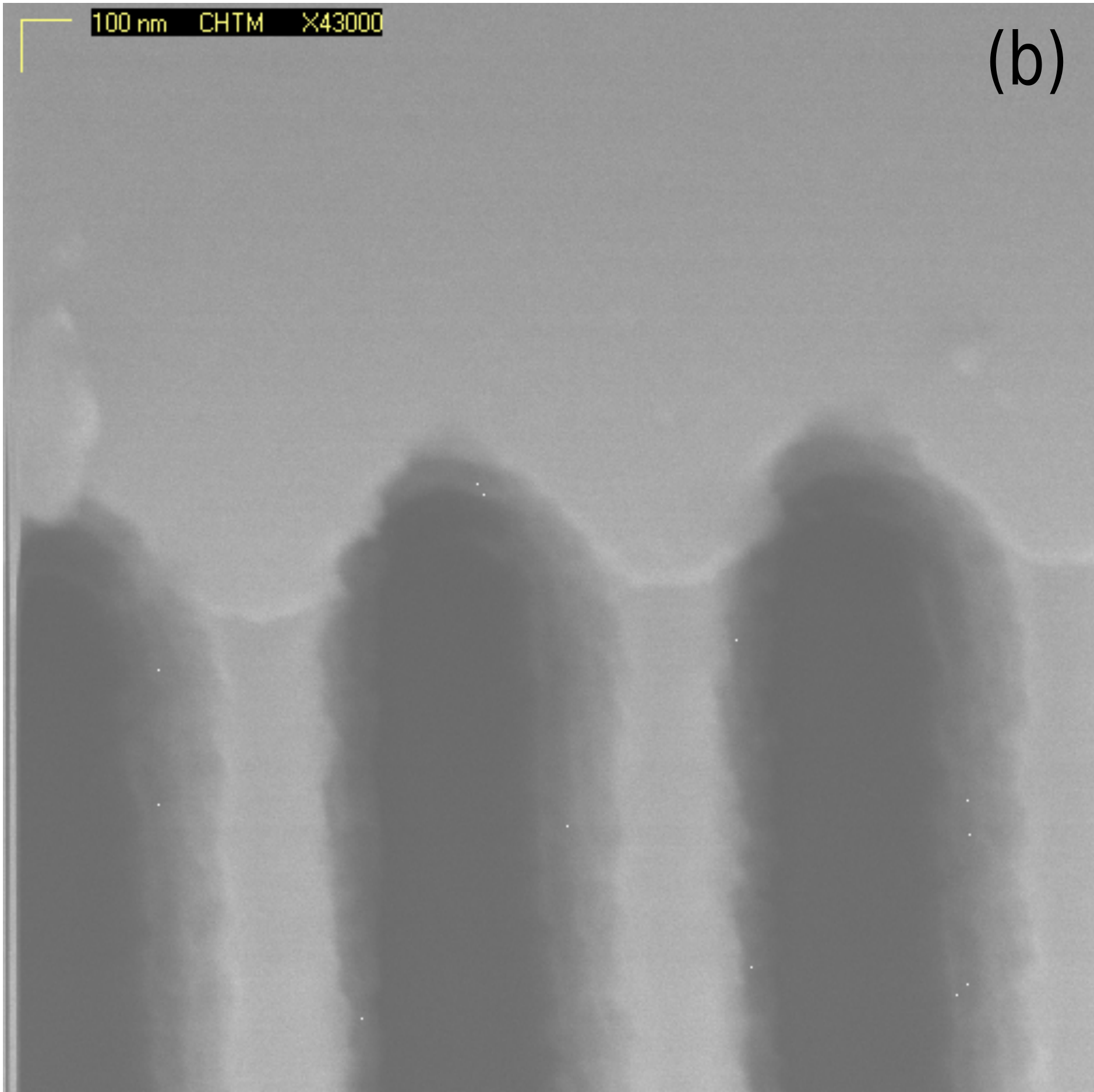}
\includegraphics[width=0.3\textwidth]{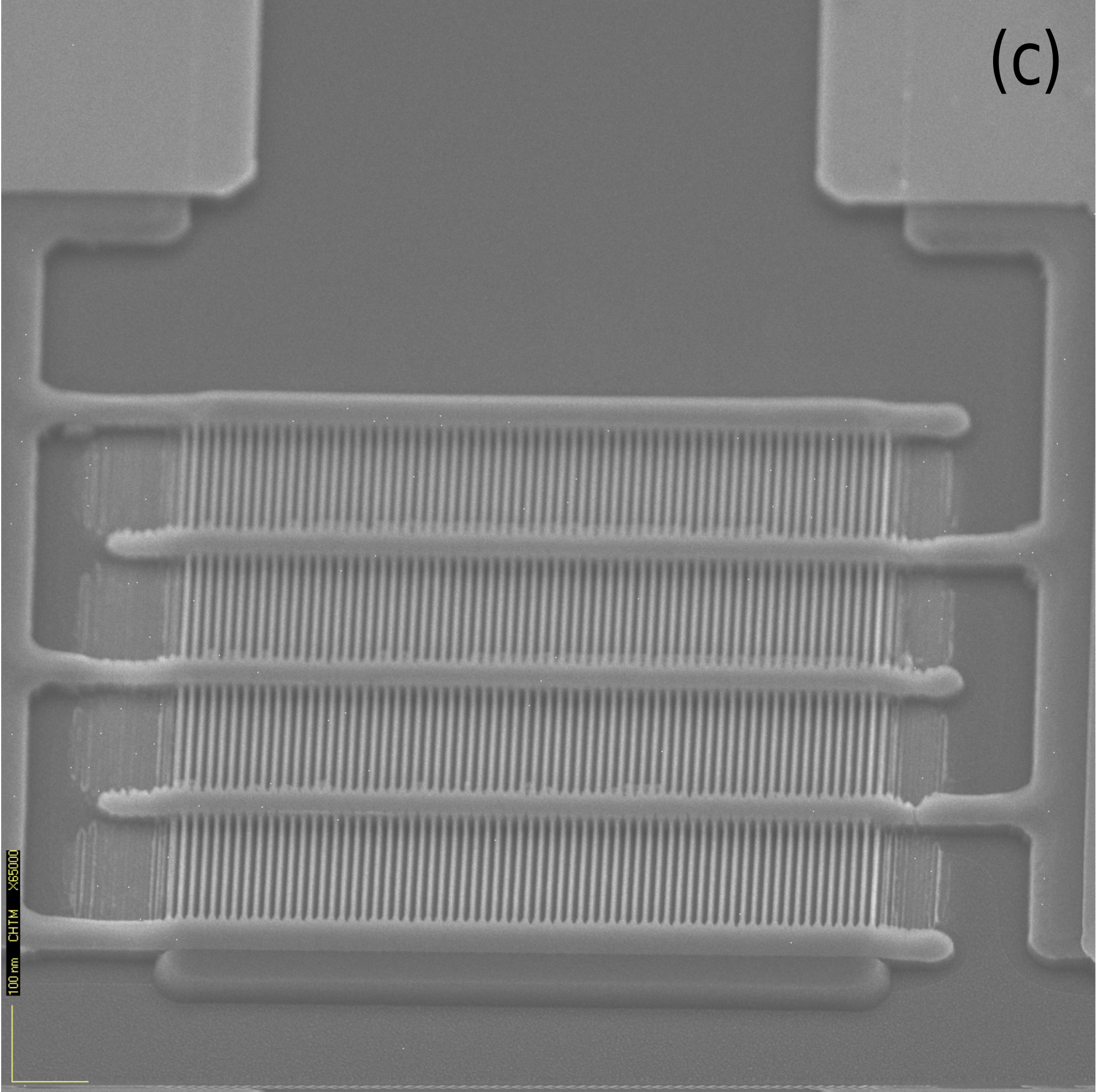}
\caption{\label{f3}
SEM image of (a) pre-oxidized Si mesa configuration with precursors to wall structures in the active region in-between planar un-textured regions where the metal contacts will be deposited; (b) Planar un-textured Si where thermally grown oxide was removed for metal contact deposition connecting walls; (c) Fully fabricated wall device with interdigitated electrodes.}
\end{figure}

The wall structured samples were then configured into two terminal metal-Si/nanowall-
metal (MSM) devices for optical and electrical characterization. The MSM device configuration
was specifically designed so the current would
ow within the wall boundaries between
the electrodes. This allowed the physical cross-section of the wall structures to dictate the
current flow properties. The mesa structures were fabricated to cutoff any stray current paths
that could bypass the intended active region (wall) carrier path. Figure\ \ref{f3}(a) shows a SEM
picture of a typical pre-device mesa structure. After the walls were oxidized to achieve the
desired wall width, the thermally grown oxide was selectively removed from the planar un-
textured Si pad locations [Figure\ \ref{f3}(b)] using an appropriate photo-mask and a chemical $1:6$
buffered oxide etch (BOE) process.
\medskip

Following the resist removal the samples were cleaned using a sulfuric-acid/hydrogen-peroxide solution,
and a DI water rinse followed by a nitrogen gas dry step. The samples were
then re-patterned using photoresist and a second mask was used in the process to form the
electrode contact regions. Three separate evaporations (30 nm of Ni) were performed.
The first one was performed at a normal incidence to the sample surface and the other two at a $30^{\rm o}$
degree tilt angles in order to ensure complete coverage of the mesa step height. After Ni
evaporation, liftoff was performed to remove the unwanted metal and resist using acetone.
Following a thorough clean using methanol/DI-water, the samples were again dehydrated
and spin-coated with a thick resist layer. The samples were patterned using a final
metallization mask set. A layer of Cr and Au was evaporated on the electrode regions. $30$\,nm/$200$\,nm
of Cr/Au were evaporated and liftoff process was used to remove the resist and
unwanted metal. Figure\ \ref{f3}(c) shows SEM pictures of a fully fabricated wall device.

\section{Electrical and Optical Measurements and Analysis}
\label{sec3}

\subsection{DC measurements}
\label{sec3.1}

At room temperature only a small number of carriers are thermally generated (as dark current)
for a Si bandgap of $1.15$\,eV. At low bias voltages (linear region of operation) the slope
of the $I$-$V$ dark current is proportional to the device resistance that includes contributions
of thermally generated carriers from both the wall channels and the metal/semiconductor
contact regions. At higher biases the current saturates when all thermally generated carriers
are collected. Any further increase in the current can be attributed to leakages across the
contact metal-semiconductor barrier and to non-linear generation of carriers across the barrier.\,\cite{r17}
The back interpolation of this leakage current to the zero bias ($0$\,V) is a measure of
the saturated dark current ($I_{\rm ds}$). Although the photocurrents are a few orders of magnitude
larger than the thermally generated dark currents, the analysis of the photocurrent ($I_{\rm ps}$) $IV$
function is the same as the dark current ($I_{\rm ds}$) $IV$ plots. For dc response analysis, two sets of
measurements were performed. These include: (i) dark currents as a function of wall width
thickness, and (ii) photocurrents as a function of wall width thickness. These results are discussed
and analyzed below.

\subsection{Dark currents versus wall width thickness}
\label{sec3.2}

\begin{figure}[p]
\centering
\includegraphics[width=0.45\textwidth]{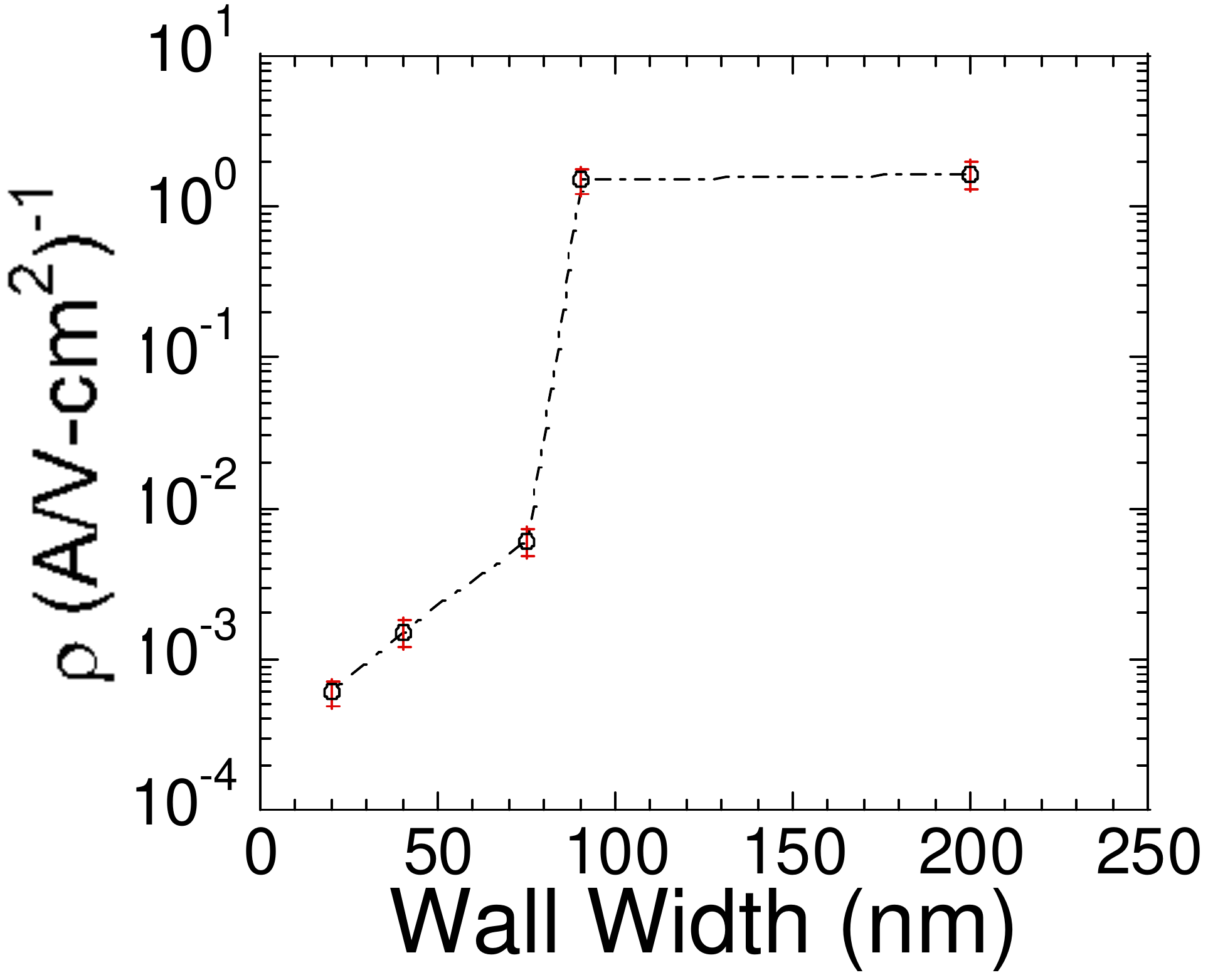}
\caption{\label{f4}
Plot showing resistivity characteristics as a function of down scaling the wall widths.}
\end{figure}

To study the carrier conduction properties versus dimensionally scaling down the width
of the wall structures into the nano-regime, the samples were characterized in batches.
Using samples with wall widths of $200$\,nm, $95$\,nm, $75$\,nm, $40$\,nm, and $20$\,nm, the room
temperature dark currents were measured with a probe station and digital $I$-$V$ curve tracer.
As the physical cross-sectional area of the wall widths was reduced from $200$\,nm to $20$\,nm,
we know from Ohm's law, the resistance should increase linearly as a function of area. In
other words the resistivity in units of
$\Omega/\square$ should remain constant. However, as can be seen
from the Figure\ \ref{f4}, the resistivity is not constant but drops significantly as the width of the
wall is reduced below $95$\,nm. This suggests that there is an increase in conductivity as the wall
thickness decreases from $95$\,nm to $20$\,nm. Since the number of thermally generated carriers
is directly proportional to the volume of the active region, any increase in the conductivity,
as wall width cross-sectional region decreases from $95$\,nm to $20$\,nm, cannot be attributed to
the volume of the semiconductor material, but must be the result of a substantial increase in
the carrier velocity. Confirmation of this hypothesized mechanism was obtained with the use
of transient time analysis as discussed in subsection\ \ref{sec3.4}.

\subsection{Photocurrents versus wall width thickness}
\label{sec3.3}

\begin{figure}[p]
\centering
\includegraphics[width=0.45\textwidth]{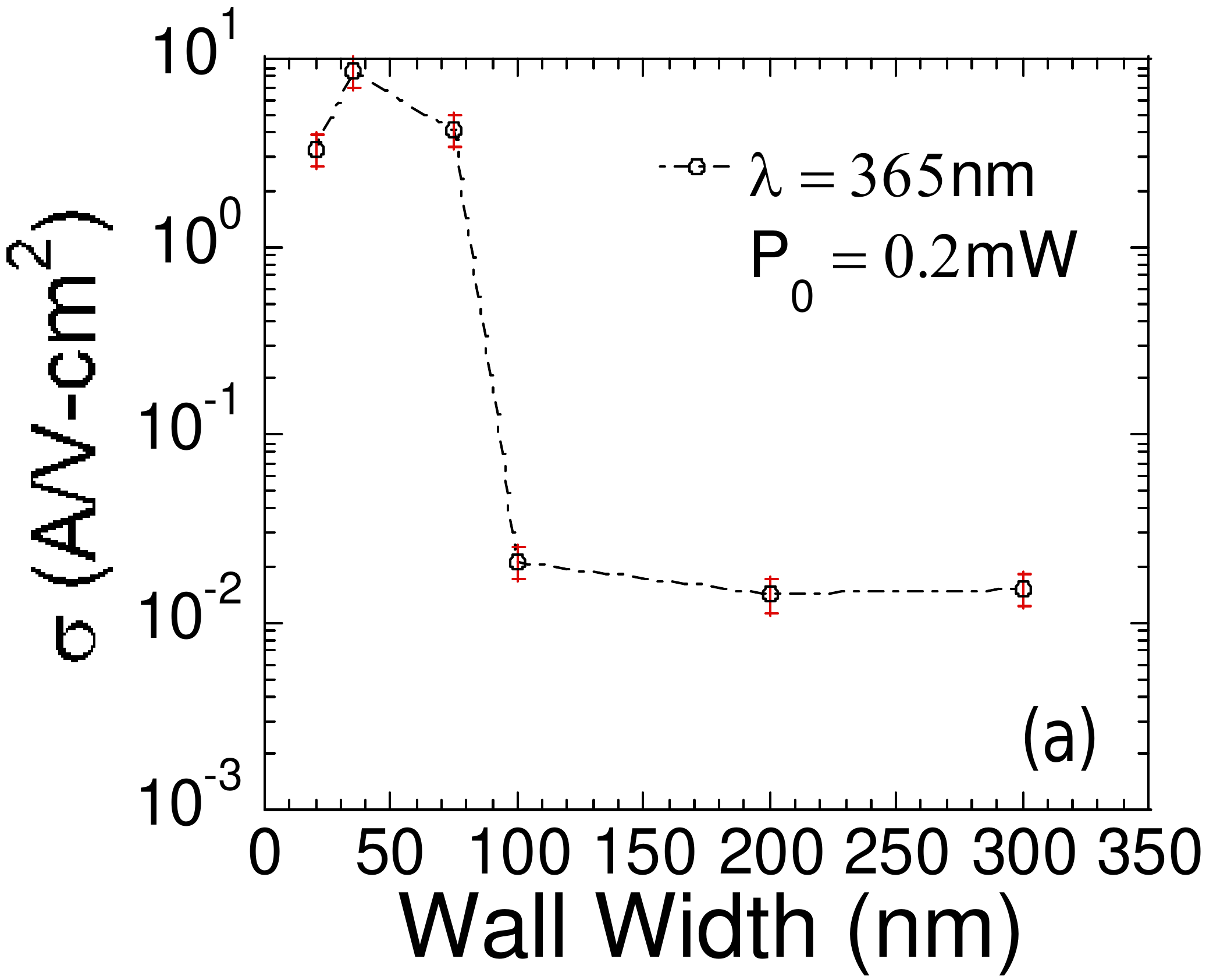}
\includegraphics[width=0.45\textwidth]{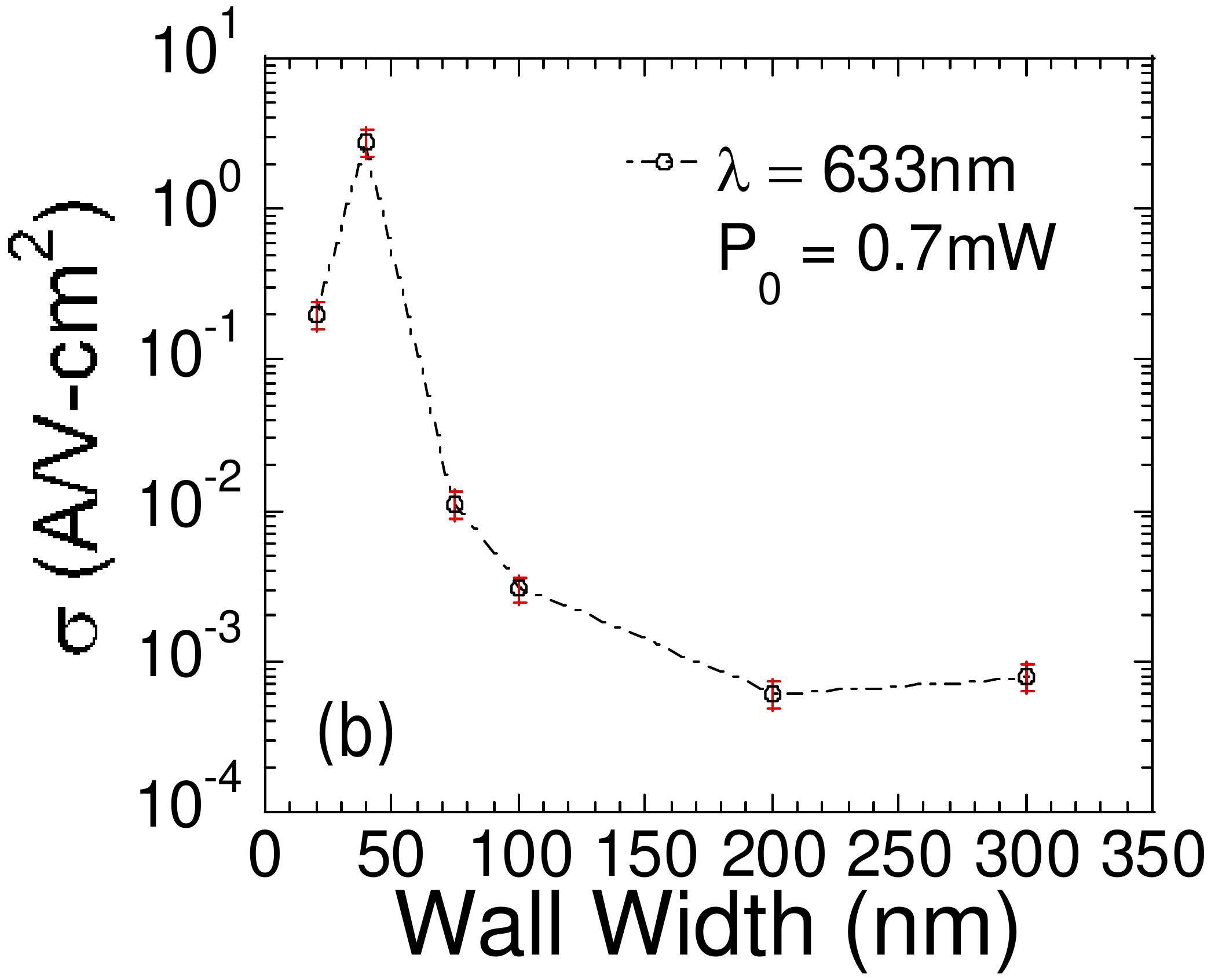}
\caption{\label{f5}
Plots of photoconductivity characteristics as a function of down scaling the wall widths (a) for $\lambda=365$\,nm; (b) for $\lambda=633$\,nm.}
\end{figure}

DC steady state photocurrents were measured using a $365$\,nm wavelength, $1.132$\,W$/cm^2$
argon-ion laser and a $633$\,nm wavelength, $3.96$\,W$/cm^2$ HeNe laser. The laser beam spot
diameter was less than $8\,\mu$m and was focused within the active region of the electrode
spacing covering several wall structures. By using $365$\,nm and $633$\,nm wavelengths, a more
complete insight into absorption and carrier transport as a function of wall thickness can be
achieved. At $365$\,nm, absorption occurs within the top first $10$\,nm of the Si wall structures
with heights of $1500$\,nm. For $633$\,nm the total photon absorption extends through the entire
wall height. Figure\ \ref{f5}(a) and \ref{f5}(b) show the conductivity versus wall thickness profiles
respectively. As can be noted from the figures, a peak in the conductivity occurs around the $40$\,nm
(physical wall width) samples followed by a decrease around $25$\,nm width samples. The
significance of this can be explained through the effects of strain inside the wall structures that
affect the carriers mobilities as the dimensions are reduced, as discussed in section\ \ref{sec4}.

\subsection{Transient time response measurements and analysis}
\label{sec3.4}

\begin{figure}[p]
\centering
\includegraphics[width=0.65\textwidth]{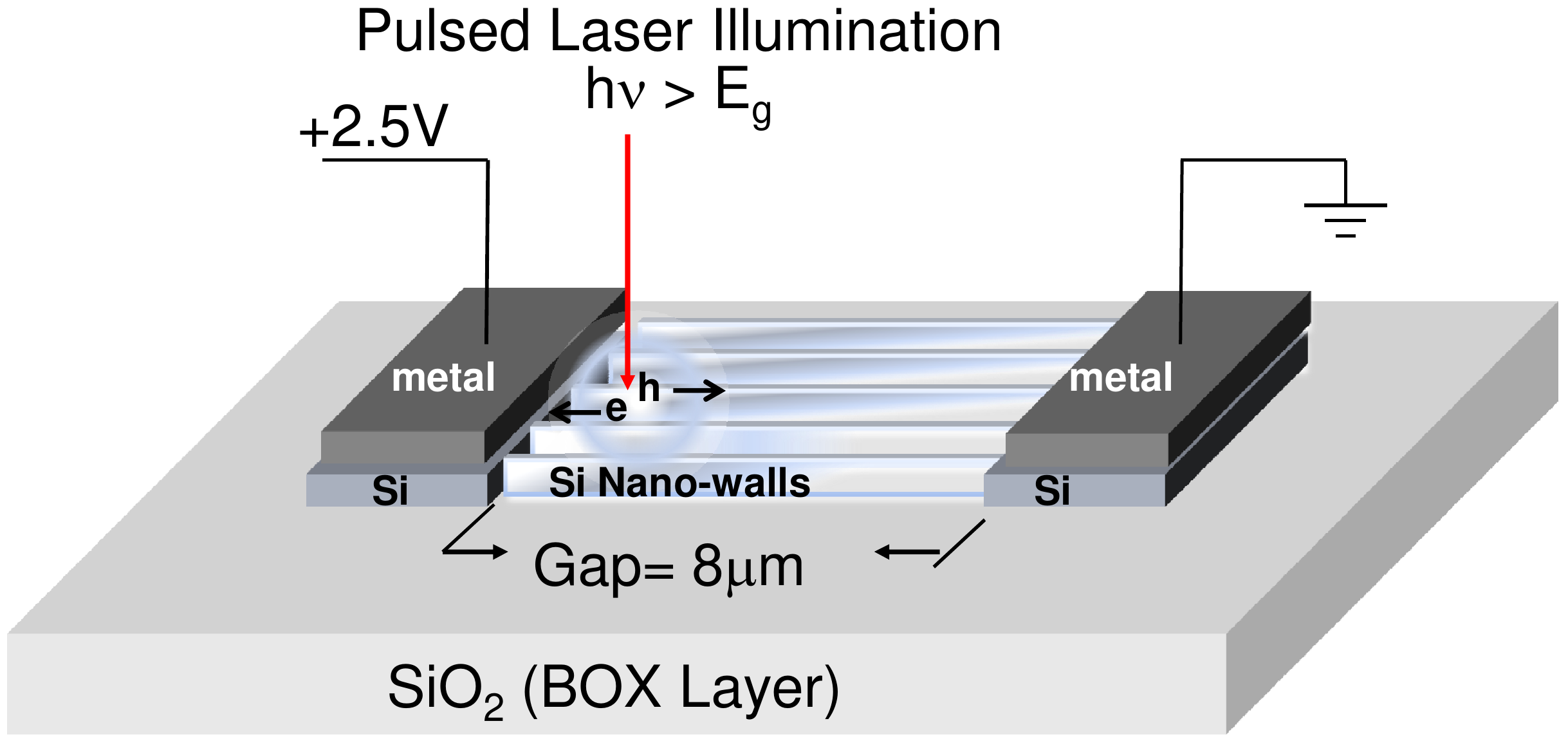}
\caption{\label{f6}
(Color online) Schematic configuration of a wall structured MSM device used for carrier time response measurements.}
\end{figure}

The schematic of the pulsed carrier transport experiment is shown in Figure\ \ref{f6}. This setup is
based on a modified version of the Haynes-Schockley\,\cite{r21} experiment. This measurement provides
an unambiguous direct measure of the actual transit time of electrons and holes through the
channel. When a narrow pulse of light strikes the wall structured active region of the device
near the left electrode as shown in Figure\ \ref{f6}, equal number of electrons and holes are
generated, and are then subjected to diffusion and drift forces in a presence of an electric field.
Based on the experimental configuration the electrons will be rapidly collected near the
positively biased electrode and the holes will have to travel the entire channel to the
negatively biased electrode. From the measured time response signal profile at the opposite
electrode, the hole transient time limited carrier velocity can be determined, provided the
carrier lifetime is greater than the total transit time. If the optical pulse of light strikes near
the opposite electrode, the holes will be rapidly collected and the electrons would have to
transit through the channel, thus the measured signal at the opposite electrode would be
electron transit time limited.
\medskip

The pulsed response measurements were taken using a $150$-fs duration excitation at $\lambda=400$\,nm
from a cw mode-locked Ti:Al$_2$O$_3$ laser (doubled for the short wavelength, $0.2$\,mW average
power at a $77$\,MHz repetition rate).
\medskip

The wall structured MSM devices were probe tested using an $18$\,GHz probe and a high-speed
digital sampling oscilloscope with an approximately $1$\,ps resolution capability. The laser spot
size was $1\,\mu$m in diameter and the electrode gaps were $8\,\mu$m. Normal incidence was used for
the experiment. The time response measurements were taken for low electric field strengths $~3\times 10^3$V$/$cm,
($2.5$\,V across $8\,\mu$m gap) thus avoiding velocity saturation.
\medskip

Before the experimental data and analysis is provided it is useful to review the three
primary factors that can impact the carrier transport through a semiconductor region. These factors are:\\
\begin{itemize}
\item Field dependent velocity of carriers through the active region. At high $E$-fields, the
velocities of both electrons and holes in Si saturate at about $1\times 10^7$\,cm$/$s,\\,\cite{add19} provided
the field within the electrodes exceeds the saturation value for most of its length, we can
assume that the carriers move with a average velocity drift. Velocity saturation is not
an issue in our experiment since the applied field is much lower than what is required for saturation.

\item Diffusion of carriers in the active region. The time it takes for carriers to diffuse a
distance $d$ is $\tau_{\rm diff}=d^2/2D$ where $D$ is the carrier diffusion coefficient. The diffusion of
carriers becomes a two dimensional process as the thickness of the Si wall-structures is
reduced and carriers are physically constricted in movement by the Si/SiO$_2$ interfaces from all sides.

\item Junction and parasitic capacitance effects. A metal-semiconductor junction under
reverse bias exhibits a voltage-dependent capacitance caused by the variation in stored
charge at the junction represented by the relation $C_{\rm J}=({\cal A}/2)\,\sqrt{2e\epsilon_0\epsilon_{\rm r}N_{\rm d}/V}$,
where ${\cal A}$ is the junction cross-sectional area, $N_{\rm d}$ is the ionized donor density, $\epsilon_{\rm r}$ is the dielectric
constant and $V$ is the junction voltage. This capacitance is usually quite small for
MSM device structures as a result of their planar electrode design. There are also
parasitic circuit capacitances associated with the probing and cabling that usually
dominate the electrical response as well as the limiting response of the electronics. For
this study all film devices have an identical circuit limitation.
\end{itemize}
\medskip

\begin{figure}[p]
\centering
\includegraphics[width=0.65\textwidth]{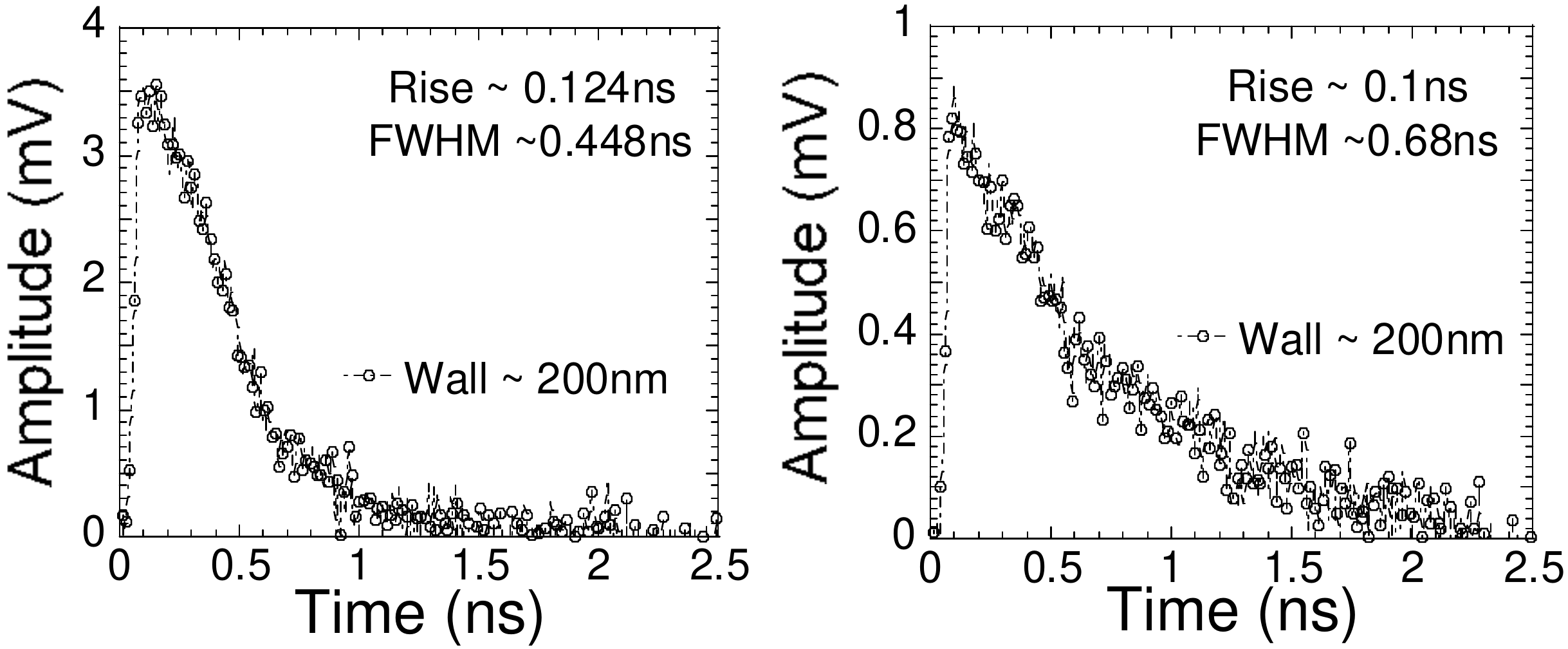}
\includegraphics[width=0.65\textwidth]{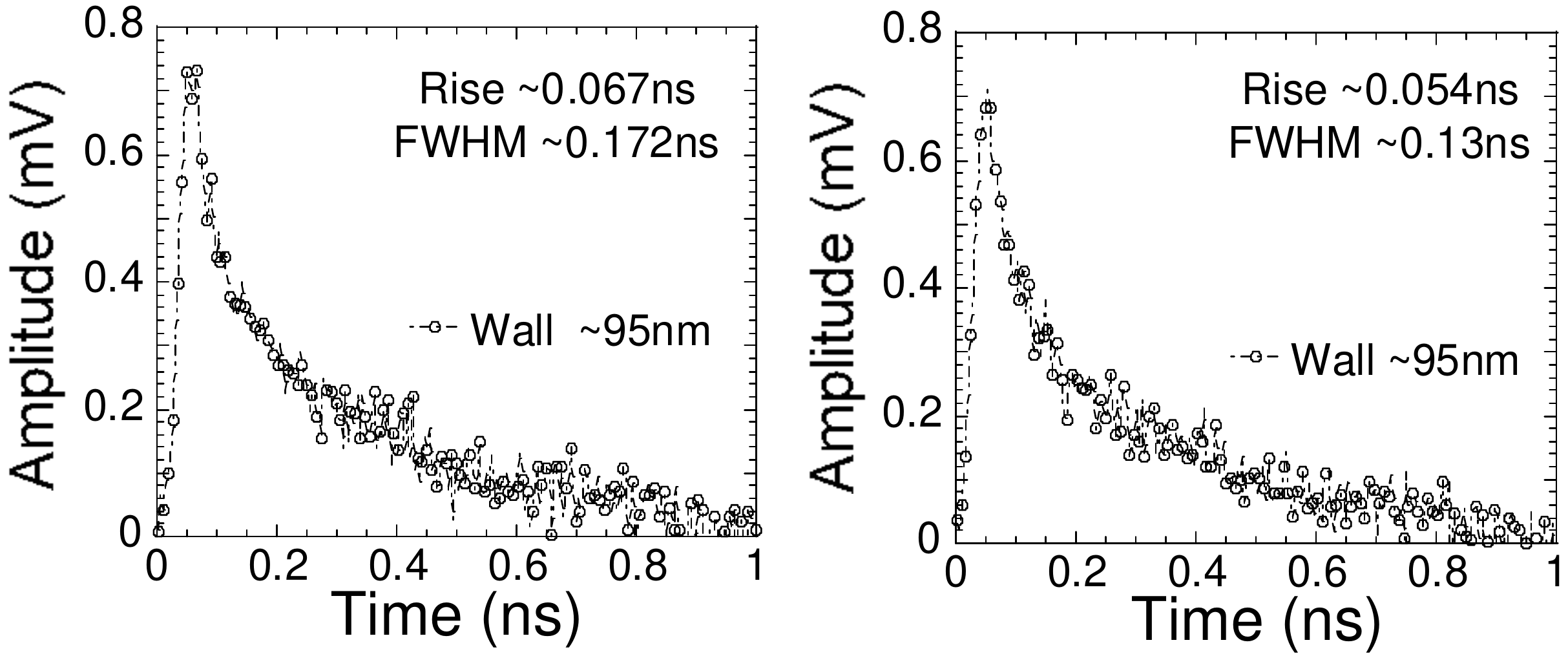}
\includegraphics[width=0.65\textwidth]{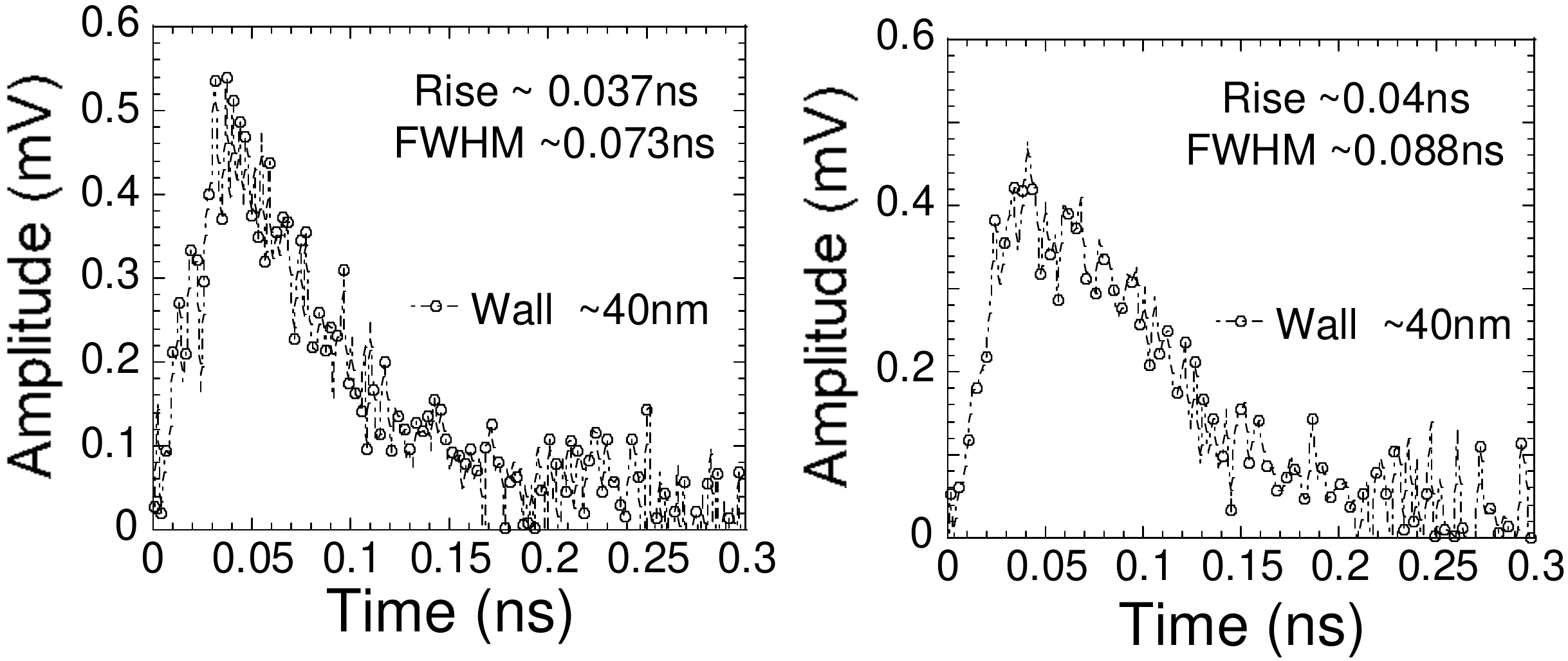}
\includegraphics[width=0.65\textwidth]{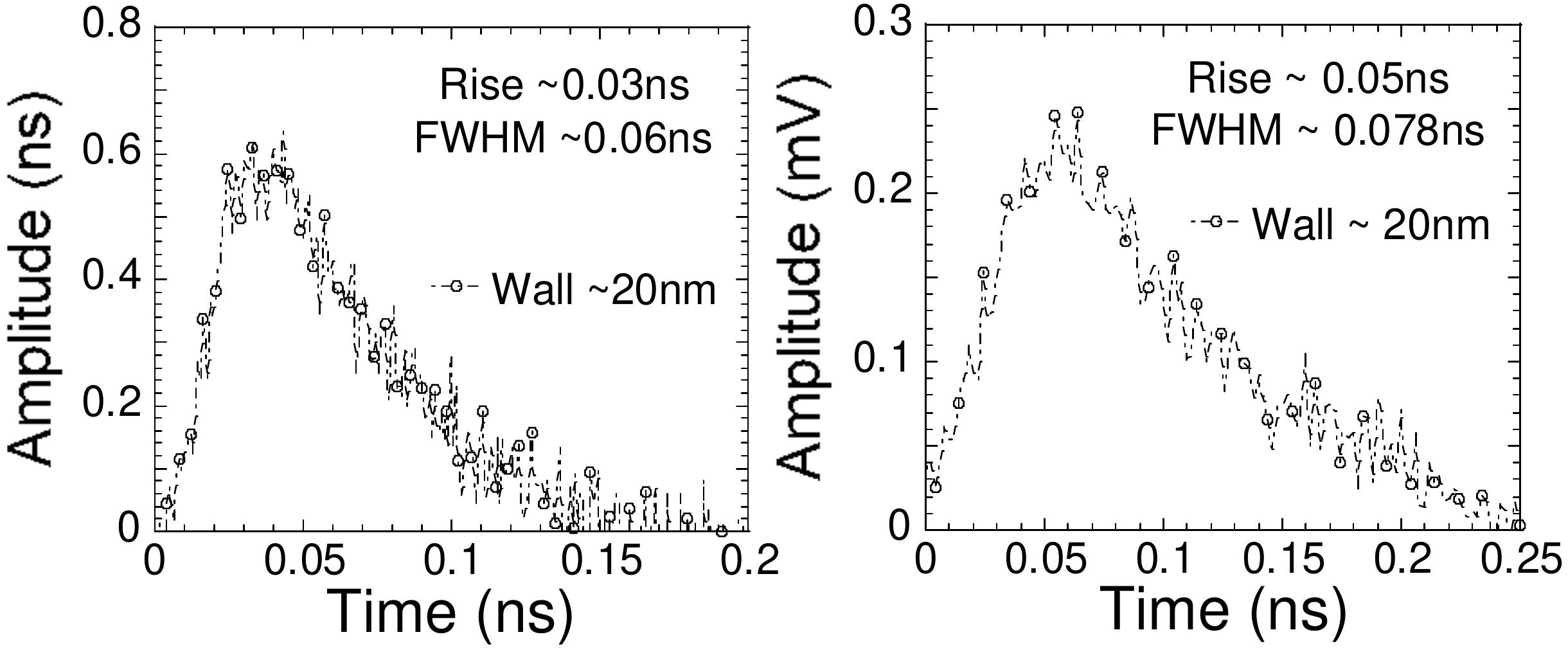}
\caption{\label{f7}
Measured time response signals of $~200$\,nm wall (row-1), $~95$\,nm wall (row-2), $~40$\,nm wall (row-3), and $~20$\,nm wall (row-4).}
\end{figure}

Figure\ \ref{f6} shows the bias polarity of our experiment in which the left
electrode polarity is positive and the right electrode is ground. With this bias configuration
once a pulse of light with a spot size $<1\,\mu$m, as in the case of our experiment, strikes within
the active region, the holes travel towards the right electrode and the electrons travel in the
opposite direction towards the left electrode. Figure\ \ref{f7}(a)-\ref{f7}(d) shows the experimental results
of the time response measurements for $200$\,nm, $95$\,nm, $40$\,nm and $20$\,nm thick wall devices for
both electron and hole dominated signals. From a first pass, as can be seen from these plots, as
the thickness of the wall-channels are decreased, the time response signal decays faster. In
particular in the case of the $40$\,nm and $20$\,nm thick walls the signal decays over an order of
magnitude faster then the $200$\,nm sample for both electrons and holes. The rise time of the
signals is an important parameter, since it directly provides the carrier transit time.\,\cite{grundmann} The rise
time ($t_{\rm d}$) is defined as is the time-lapse from moment when the pulse of light strikes one end of
the active region of the MSM, near one electrode, and the moment when the photo-generated
carrier signal is detected at the opposite electrode. From the rise time data, provided on Fig.\,\ref{f7}(a)-\ref{f7}(d),
we can determine the carrier mobilities as a function of wall thickness as follows.\,\cite{grundmann}
\medskip

First from the experimental time response measurements we can calculate the average
carrier velocities by applying the given relation,

\begin{equation}
v_{\rm Carrier-Velocity}=\frac{({\rm Electrode\ gap})}{t_{\rm d}}\ \ \mbox{(cm/s)}\ ,
\label{e1}
\end{equation}
where $t_{\rm d}$ is the average time it takes for the pulsed carrier signal to cross the electrode
gap distance. The pulse travels in the presence of a field and expands from its originating
point due to diffusion. In this case we are ignoring the RC time delay that the pulsed
signal experiences once it reaches the edge of the depletion region near the electrodes since
the widths of the depletion regions are very small in the sub-micron range compared to the
electrode gap which is $8\,\mu$m in length.
\medskip

By definition the average carrier mobility can be written as,

\begin{equation}
\mu_{\rm avg}=\frac{v_{\rm Carrier-Velocity}}{V_{\rm bias}/({\rm Electrode\ gap})}\ ,
\label{e2}
\end{equation}
where $V_{\rm bias}$ is the external bias applied to the electrodes.
\medskip

\begin{figure}[p]
\centering
\includegraphics[width=0.45\textwidth]{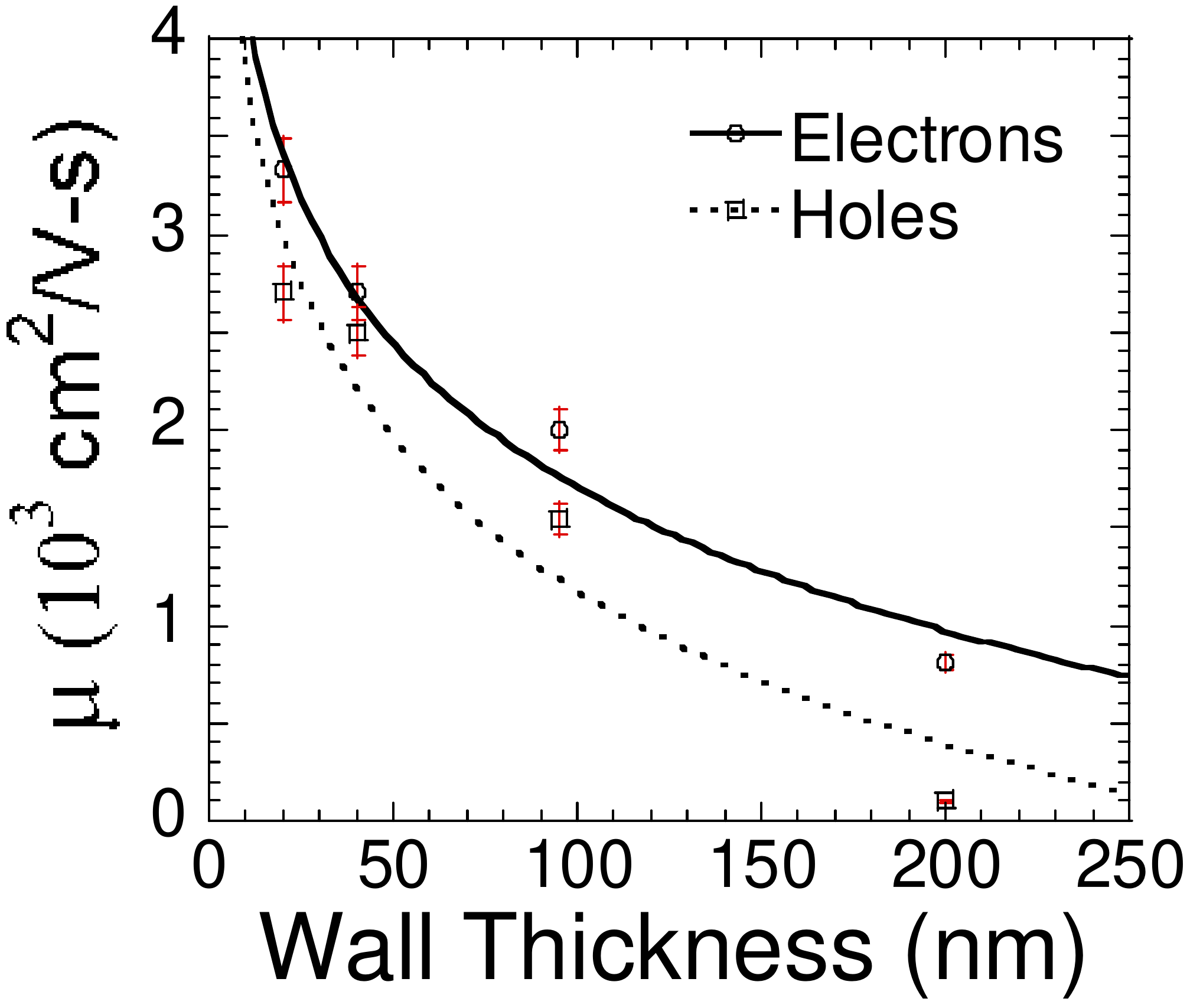}
\caption{\label{f8}
Carrier mobility values calculated from direct measure of rise time values as a function of wall thickness.}
\end{figure}

Figure\ \ref{f8} shows a plot of average field dependent electron and hole limited mobility values
using experimental values of rise time, $t_{\rm d}$, and the above expression as a function of wall
thickness. We know that the carrier transport of electrons and holes in the thickest wall
sample ($200$\,nm) is essentially similar to the transport properties in bulk silicon. However
we observe a considerable increase in low field dependent mobility values below $75$\,nm wall thicknesses. Recall the fact that we actually have a much narrower effective cross-sectional
regions from which carriers propagate due to the repulsive nature of the boundary at the Si/SiO$_2$ interface, and the carrier profile tends to peak a certain distance away from the
interface close to the center of the wall structures.\,\cite{r10} At these nanoscales we must account for
the strain effects, which include the reversal splitting of light- and heavy- hole bands as
well as the decrease of conduction-band effective mass by reduced Si bandgap energy. These
strain effects are formulated in our microscopic model for explaining the experimentally observed
enhancements in both conduction- and valence-band mobilities with reduced Si wall
thickness, {\em i.e.\/} consider the case where the hole mobility is given by $\mu_{\rm h}=e\tau_{\rm h}/m_{\rm h}^\ast$, where
$1/m_{ij}^\ast=(1/\hbar^2)\,(\partial^2E(k)/\partial k_i\partial k_j)$.
The narrower light-hole band dominating the transport
can have a significant enhancement on the overall mobility which is consistent with our
experimental result. Specifically, the enhancements of the valence-band and conduction-band
mobilities are found to be associated with different aspects of physical mechanisms. The role
of the biaxial strain buffering depth is elucidated and its importance to the scaling relations
of wall-thickness is reproduced theoretically. A detailed theoretical model is described in the
next section which explains our experimental results in a comprehensive manner.

\section{Strain Effects Modeling to Explain the Rise in Electron and Hole Mobility}
\label{sec4}

\begin{figure}[p]
\centering
\includegraphics[width=0.85\textwidth]{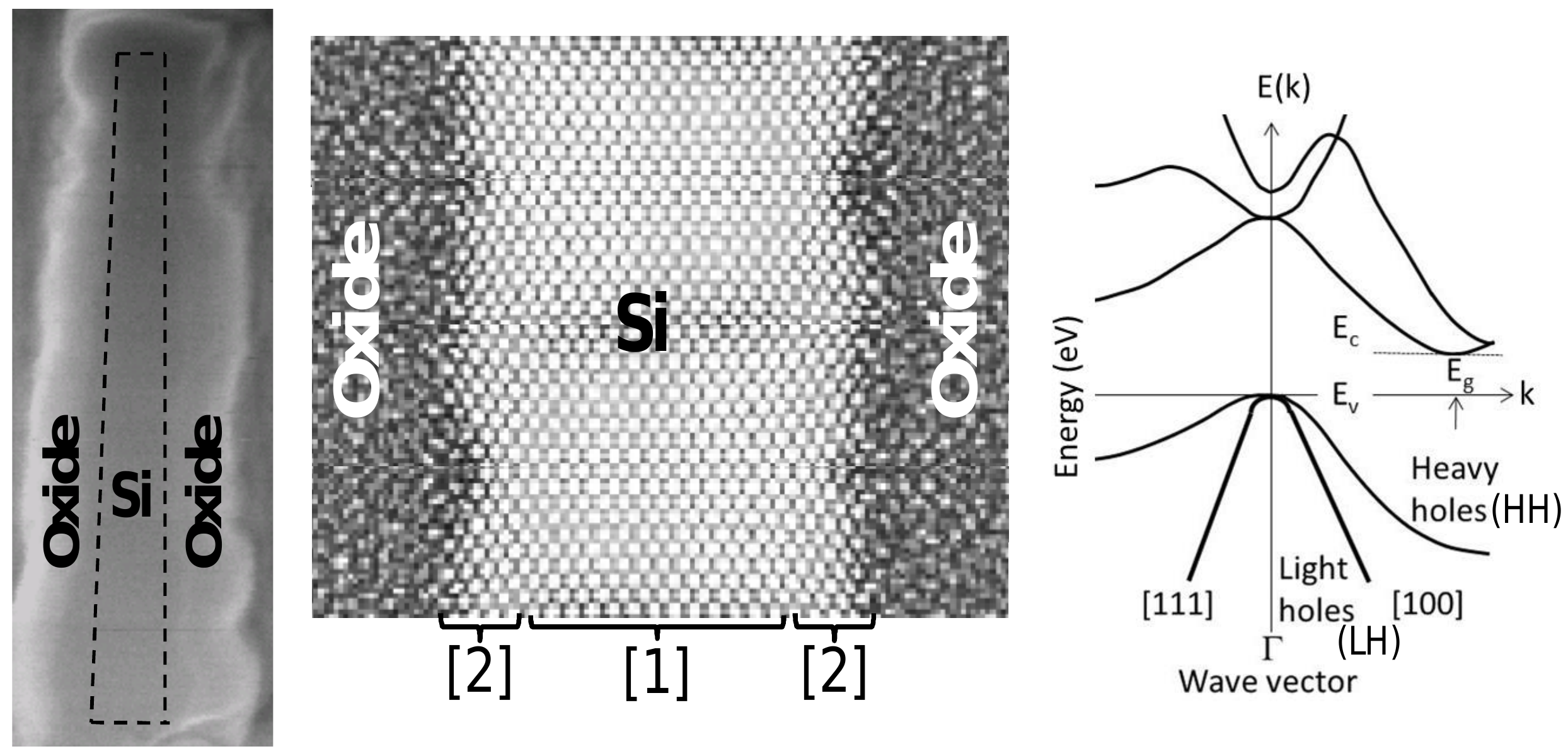}
\caption{\label{f9}
SEM image of single $~200$\,nm wall (left) and artists depiction of Si and O atoms shown by light and dark gray spheres, respectively (middle). In the middle panel, region-[1] is unstrained while region-[2] is strained. In thickest structures strained region is near the interfaces, but due to fixed oxide charges the current flows away in the unstrained region. The right panel shows $E$-$k$ band diagram of unstrained region-[1].}
\end{figure}

\begin{figure}[p]
\centering
\includegraphics[width=0.85\textwidth]{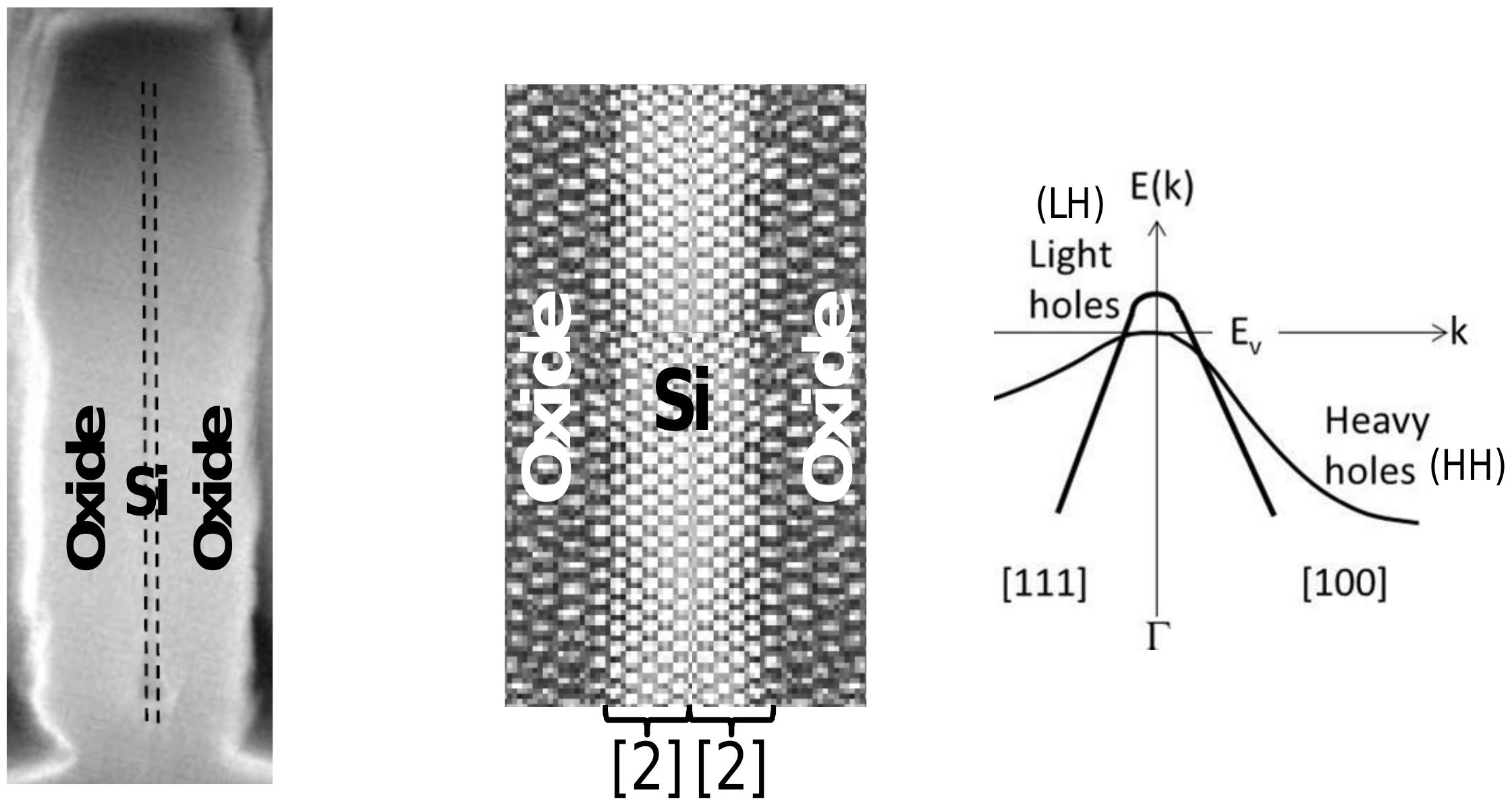}
\caption{\label{f10}
SEM image of single $~20$\,nm wall (left) and artists depiction of Si and O atoms shown by light and dark gray spheres, respectively (middle). Note unstrained region-[1] in the middle panel has vanished as strained region-[2] closed in from both sides. In thinnest structures the strain is continuous throughout the wall. The right panel displays $E$-$k$ band diagram of strained region-[2].}
\end{figure}

Figures\ \ref{f9}(a)-\ref{f9}(c) represent the thickest wall channels and Figures\ \ref{f10}(a)-\ref{f10}(c) represent the
thinnest wall channels. Note that the associated E-k diagrams of Fig.\,\ref{f9}(c) and Fig.\,\ref{f10}(c) represent
the center regions of the wall channel structures where the carriers flow through.
\medskip

If we consider a total valence-band hole concentration $n_{\rm v}$ then, the light-hole ($n_{\rm LH}$) and the heavy-hole ($n_{\rm HH}$) concentration will satisfy the charge-conservation relation
$n_{\rm LH}+n_{\rm HH}=n_{\rm v}$, where

\[
n_\sigma=\frac{g_{\Gamma}g_s}{{\cal V}}\sum_{{\bf k}}\,\left[1+\exp\left(\frac{E_k^\sigma\mp\Delta E_{\rm str}^{\rm v}-u_{\rm v}}{k_BT}\right)\right]^{-1}
\]
\begin{equation}
\approx 2g_{\Gamma}\left(\frac{m^\ast_\sigma k_BT}{2\pi\hbar^2}\right)^{3/2}\,\exp\left(\frac{u_{\rm v}\pm\Delta E^{\rm v}_{\rm str}}{k_BT}\right)\ .
\label{e3}
\end{equation}
Here, the subscript $\sigma$ takes HH or LH and the upper (lower) sign corresponds to HH (LH) state.
In the above expressions, the approximations are made for high temperatures, ${\cal V}$ is the volume of the silicon film, $T$ is the system temperature, the zero energy is chosen at the middle point between the split
pair of light-hole and heavy-hole bands, ${\bf k}$ is the three dimensional wave vector of carriers, $g_\Gamma=2$ (not $6$ due to strain effect) is the $\Gamma$-valley degeneracy for holes and $g_s=2$ is the spin degeneracy
for both light-holes and heavy-holes. In addition, $u_{\rm v}$, which depends on both $T$ and $n_{\rm v}$, is the chemical potential to be determined for valence bands, $E_k^{\rm HH}=\hbar^2k^2/2m^{\ast}_{\rm HH}$ is the
kinetic energy of heavy holes and $E_k^{\rm LH}=\hbar^2k^2/2m^{\ast}_{\rm LH}$ is the kinetic energy of light holes, where $m^{\ast}_{\rm HH}=0.49\,m_0$ and $m^{\ast}_{\rm LH}=0.16\,m_0$ ($m_0$ is the free-electron mass)
are the effective masses for heavy holes and light holes, respectively. Additionally, $\Delta E_{\rm str}^{\rm v}$ introduced in the above expressions stands for the half of the valence-band splitting due to the existence
of strain.
\medskip

From Eq.\,(\ref{e3}) and $n_{\rm LH}+n_{\rm HH}=n_{\rm v}$, we obtain $n_{\rm LH}/n_{\rm v}=[1+\gamma^{3/2}\,\exp(2\Delta E_{\rm str}^{\rm v}/k_BT)]^{-1}$ and $n_{\rm HH}/n_{\rm v}=1-n_{\rm LH}/n_{\rm v}$,
where $\gamma=m_{\rm HH}^\ast/m_{\rm LH}^\ast>1$. For biaxial and shear strains\,\cite{grundmann,bahder}, we have the valence-band splitting, given by
$\Delta E_{\rm str}^{\rm v}=\pm\left\{(b^2/2)\left[(\epsilon_{xx}-\epsilon_{yy})^2+(\epsilon_{yy}-\epsilon_{zz})^2+(\epsilon_{zz}-\epsilon_{xx})^2\right]
+d^2\left[\epsilon_{xy}^2+\epsilon_{yz}^2+\epsilon_{xz}^2\right]\right\}^{1/2}$, where the upper sign is for the compressive strain while the lower sign for the tensile strain in the direction perpendicular to the
interface of silicon and silicon-dioxide materials, $b$ and $d$ are the optical deformation potentials, and $\epsilon_{jj^\prime}$ represents the strain tensor in the three dimensional space with $j,\,j^\prime=x,\,y$,
and $z$, the diagonal matrix elements $\epsilon_{jj}$ are associated with biaxial strain, and the off-diagonal matrix elements $\epsilon_{jj^\prime}$ with $j\neq j^\prime$ correspond to contributions from the shear strain.
For silicon crystals, we have $b= -2.33$\,eV and $d= -4.75$\,eV.
\medskip

If we choose the $z$ direction as the direction perpendicular to the interface for biaxial strain we simply get\,\cite{schaffler} $\epsilon_{xx}=\epsilon_{yy}=\epsilon_{\|}$, $\epsilon_{zz}=\epsilon_\perp$, and $\epsilon_{ij}=0$ for $i\neq j$, where $\epsilon_{\|}=(a_{\rm \|,\,Si}/a_{\rm Si}-1)$, $\epsilon_{\perp}=(a_{\rm \perp,\,Si}/a_{\rm Si}-1)$. Moreover, the perpendicular lattice constant $a_{\rm \perp,\,Si}$ is related to the parallel lattice constant $a_{\rm \|,\,Si}=\bar{a}_{\rm SiO_2}$ by $a_{\rm \perp,\,Si}=a_{Si}\,[1-(2c_{12}/c_{11})\,(\bar{a}_{\rm SiO_2}/a_{\rm Si}-1)]$,
where $c_{11}=16.75\times 10^{10}$\,N$/m^2$, and $c_{12}=6.5\times 10^{10}$\,N$/m^2$ are the elastic constants of silicon. For silicon and silicon-dioxide, we have $\bar{a}_{\rm SiO_2}=(2\times4.914+5.405)/3=5.078$\,\AA\ and $a_{\rm Si}=5.431$\,\AA\ for amorphous silicon-dioxide materials. Therefore, we obtain $a_{\rm \perp,\,Si}/a_{Si}=1.050$. This leads to $\epsilon_\|=-0.065$ (compressive), $\epsilon_\perp=0.05$ (tensile), and
$2\epsilon_\|+\epsilon_\perp=-0.08$.
\medskip

\begin{figure}[p]
\centering
\includegraphics[width=0.85\textwidth]{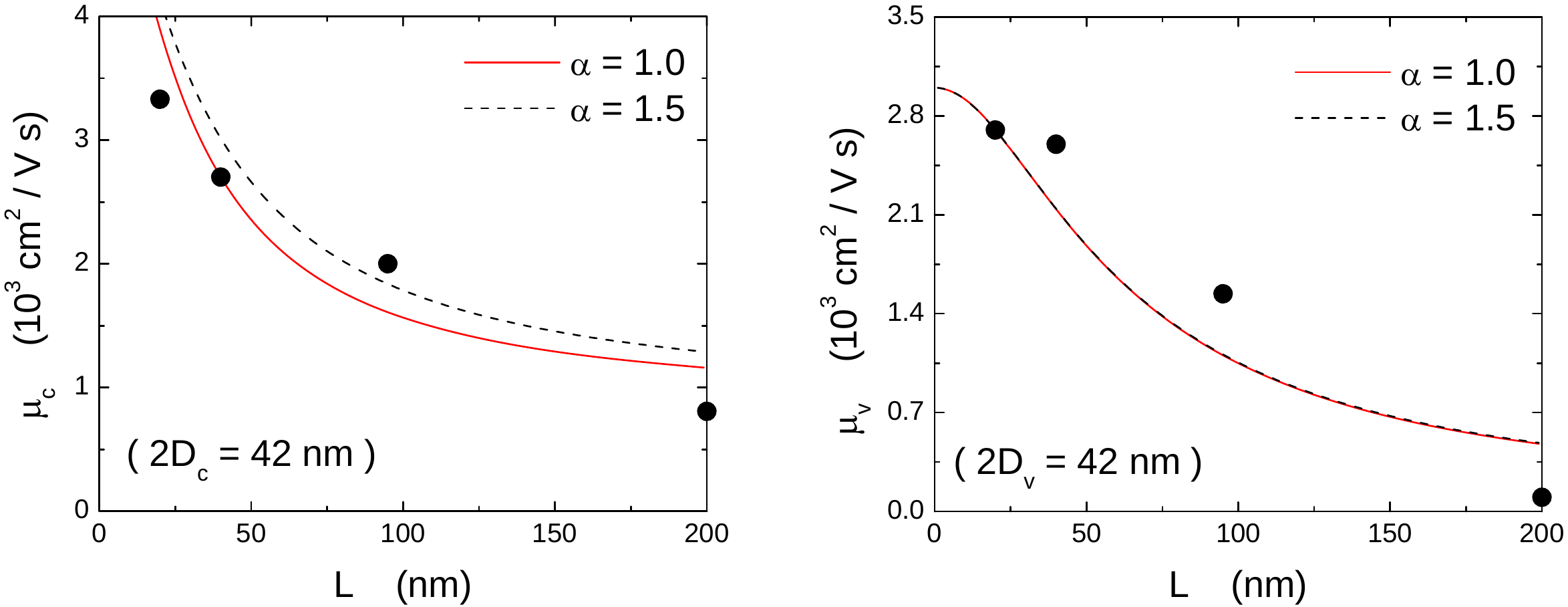}
\caption{\label{f11}
(Color Online) Theoretical modeling for electron (left panel) and hole (right panel) mobilities as functions of film thickness $L$ with $\alpha=1.0$ (red solid curves) and $1.5$ (black dashed curves)
and their comparisons with experimental data (black dots) in both panels.}
\end{figure}

The total mobility $\mu_{\rm v}$ for holes can be expressed as\,\cite{sun}

\begin{equation}
\frac{\mu_{\rm v}}{\mu_{\rm v}^{(0)}}\approx\eta_{\rm v}\,{\cal F}_{\rm v}\left(\frac{L}{\lambda_{\rm v}}\right)\,\frac{(1+\gamma^{3/2})(\gamma^\alpha+\gamma^{1/2}\,e^{{\cal A}})}
{(1+\gamma^{3/2}\,e^{{\cal A}})(\gamma^\alpha+\gamma^{1/2})}+\left(1-\eta_{\rm v}\right)\ ,
\label{e4}
\end{equation}
where ${\cal A}=2\Delta E_{\rm str}^{\rm v}/k_BT$ and $\tau_{\rm LH}/\tau_{\rm HH}=\gamma^\alpha$ (for details of calculating hole scattering time,
see Appendix\ \ref{a1}).
${\cal F}_{\rm v}(L/\lambda_{\rm v})=1+({\cal Q}_{\rm v}-1)/\sqrt{1+(L/\lambda_{\rm v})^2}$ comes from the mobility saturation effect, $\lambda_{\rm v}\sim\sqrt{3\pi^2\hbar^2/2m^\ast_{\rm LH}k_BT}$ is the quasi-quantum confinement width,
$\mu_{\rm v}^{(0)}=[(e\tau_{\rm LH}/m^\ast_{\rm LH})+\gamma^{3/2}\,(e\tau_{\rm HH}/m^\ast_{\rm HH})]/(1+\gamma^{3/2})$ corresponds to the hole mobility in the absence of strain for $L/\lambda_{\rm v}\gg 1$,
$\tau_{\rm LH}$ and $\tau_{\rm HH}$ are the scattering times for light holes and heavy holes, respectively.
Moreover, the factor ${\cal Q}_{\rm v}$ introduced in
the definition of ${\cal F}_{\rm v}(L/\lambda_{\rm v})$ is given by ${\cal Q}_{\rm v}=(\mu_{\rm v}^{\rm max}/\mu_{\rm v}^{(0)})(1+\gamma^{3/2}\,e^{{\cal A}})(\gamma^\alpha+\gamma^{1/2})/[(1+\gamma^{3/2})(\gamma^\alpha+\gamma^{1/2}\,e^{{\cal A}})]$, where $\mu_{\rm v}^{\rm max}$ is the maximum of the hole mobility in the limit of $L/\lambda_{\rm v}\to 0$.
It is clear that $\mu_{\rm v}$ increases with decreasing $L$ for the tensile strain ($E_{\rm str}^{\rm v}<0$)
in the direction perpendicular to the interface of silicon and silicon-dioxide materials, as observed by us in Fig.\,\ref{f11}.
\medskip

The values of $\eta_{\rm v}$ introduced in Eq.\,(\ref{e4})
can be scaled as $\eta_{\rm v}=1/\sqrt{1+(L/2{\cal D}_{\rm v})^2}$, where $L$ is the film thickness and $2{\cal D}_{\rm v}/L$ represents the average spatially-dependent strain
due to lattice mismatch between embedded Si crystal and surrounding amorphous SiO$_2$ material at their interface, and $L-2{\cal D}_{\rm v}>0$ represents the film effective thickness for unstrain part\,\cite{r25}. The scale of interest for these calculations of the effects of strain
near a Si/SiO$_2$ interface of a silicon nanowire was studied using molecular dynamics by Ohta, {\em et. al.\/}\,\cite{r26}. In this study, strain was most pronounced within $1$-$2$ nanometers of the interface, tensile in
the $[001]$ direction (perpendicular to the substrate) and compressive in the $[110]$ direction parallel to the substrate resulting in form of biaxial strain.
\medskip

For a given conduction-band electron concentration $n_{\rm c}$, the electron chemical potential $u_{\rm c}$, which depends on both $T$ and $n_{\rm c}$, is decided from

\[
n_{\rm c}=\sum_{\xi=X,\,L}\,n_{\xi}=\frac{g_s}{{\cal V}}\sum_{\xi=X,\,L}g_{\xi}\sum_{{\bf k}}\,\left[1+\exp\left(\frac{E^\xi_k+E^\xi_{\rm G}-u_{\rm c}}{k_BT}\right)\right]^{-1}
\]
\begin{equation}
\approx 2\sum_{\xi=X,\,L}\,g_{\xi}\left(\frac{m^\ast_{\xi}k_BT}{2\pi\hbar^2}\right)^{3/2}\,\exp\left(\frac{u_{\rm c}-E^\xi_{\rm G}}{k_BT}\right)\ ,
\label{e5}
\end{equation}
where the high-temperature approximation is made in the above expression,
$E^\xi_{\rm G}=\varepsilon^\xi_{\rm G}(T)+\Delta E^\xi_{\rm G}$ is the bandgap energy of strained silicon crystals, which depends on $T$ and the hydrostatic part of the strain,
$\varepsilon^\xi_{\rm G}$ stands for the bandgap energy of unstrained silicon crystals,
$g_{X,\,L}=2$ (not $6$ due to strain effect) represents the $X$ (in $<100>$ direction) or $L$ (in $<111>$ direction) valley degeneracy for electrons at the two minima of conduction band, $E^\xi_k=\hbar^2k^2/2m^{\ast}_{\xi}$ is
the kinetic energy of electrons and $m^\ast_{\xi}$ is the transverse effective mass of conduction-band electrons with $m^\ast_X=0.19\,m_0$ and $m^\ast_L=0.1\,m_0$.
The $T$ dependence of $\varepsilon^\xi_{\rm G}(T)$ (based on the Bose-Einstein phonon model) is given by\,\cite{grundmann}
$\varepsilon^\xi_{\rm G}(T)=\varepsilon^\xi_{\rm G}(0)-2\alpha_B\Theta_B\,[\coth(\Theta_B/2T)-1]$,
where $\alpha_B=2.82\times 10^{-4}$\,eV$/$K is a coupling constant, $k_B\Theta_B$ is a typical phonon energy with $\Theta_B=351$\,K, $\varepsilon^{\rm X}_{\rm G}(T)=1.12$\,eV and $\varepsilon^{\rm L}_{\rm G}(T)=2.4$\,eV
at $T=300$\,K for the X and L valleys.
Moreover, the strain part of the bandgap energy $\Delta E^\xi_{\rm G}$ is calculated as\,\cite{grundmann}
$\Delta E^\xi_{\rm G}=\Xi_d^{(\xi)}\,Tr(\tensor{\epsilon})+\Xi_u^{(\xi)}\,\vec{e}_\xi\cdot\tensor{\epsilon}\cdot\vec{e}_\xi+a\,Tr(\tensor{\epsilon})$,
where $\Xi_d^{(X,\,L)}$ and $\Xi_u^{(X,\,L)}$ are the deformation potentials of the conduction band for an indirect-gap silicon crystal
($\Xi_d^{(X)}=1.1$\,eV, $\Xi_u^{(X)}=10.5$\,eV for the X valley and $\Xi_d^{(L)}=-7.0$\,eV, $\Xi_u^{(L)}=18.0$\,eV for the L valley),
$a=2.1$\,eV is the difference of the deformation potentials of conduction and valence bands at two different valleys due to hydrostatic component of the strain for the silicon crystal,
and $\vec{e}_\xi$ is the unit vector pointing to the specific X or L valley.
It is clear from the above equation that $\Delta E^\xi_{\rm G}<0$ for the tensile strain and $\xi=$X or L.
\medskip

The change in the bandgap energy by strain also affects the effective mass of conduction band, given by\,\cite{cardona}

\begin{equation}
\Delta\left(\frac{m_0}{m^\ast_{\xi}}\right)\approx -\frac{E_P(2\epsilon_{\|}+\epsilon_{\perp})/3}{\varepsilon^\xi_{\rm G}(T)+\Delta_0/3}\left[2+\frac{3a}{\varepsilon^\xi_{\rm G}(T)+\Delta_0/3}\right]\ ,
\label{e6}
\end{equation}
where we have neglected the shear strain and assumed a weak strain with $|2\epsilon_{\|}+\epsilon_{\perp}|\ll 1$, $\Delta_0=44$\,meV is the spin-orbit splitting and $E_P=21.6$\,eV is the Kane energy parameter.
\medskip

The total mobility $\mu_{\rm c}$ of conduction-band electrons is obtained as

\begin{equation}
\frac{\mu_{\rm c}}{\mu_{\rm c}^{(0)}}\approx\eta_{\rm c}\,{\cal F}_{\rm c}\left(\frac{L}{\lambda_{\rm c}}\right)\,
\frac{\left[m_0/m^\ast_{X}+\Delta(m_0/m^\ast_{X})\right]^{1+\alpha}}{(m_0/m^\ast_{X})^{1+\alpha}}+\left(1-\eta_{\rm c}\right)\ .
\label{e7}
\end{equation}
where ${\cal F}_{\rm c}(L/\lambda_{\rm c})=1+({\cal Q}_{\rm c}-1)/\sqrt{1+(L/\lambda_{\rm c})^2}$ comes from the mobility saturation effect, $\lambda_{\rm c}\sim\sqrt{3\pi^2\hbar^2/2m^\ast_Xk_BT}$,
${\cal Q}_{\rm c}=(\mu_{\rm c}^{\rm max}/\mu_{\rm c}^{(0)})(m_0/m^\ast_{X})^{1+\alpha}/\left[m_0/m^\ast_{X}+\Delta(m_0/m^\ast_{X})\right]^{1+\alpha}$,
$\mu_{\rm c}^{(0)}=e\tau_X/m^\ast_X$ corresponds to the electron mobility in the absence of strain for $L/\lambda_{\rm c}\gg 1$,
$\tau_{X,\,L}$ represents the scattering times of conduction-band electrons at two different valleys and the high-energy $L$ valley has been assumed depopulated, and $\tau_X=\tau_X^0\,(m_0/m^\ast_X)^\alpha$  (for details of calculating electron scattering time, see Appendix\ \ref{a1}).
In addition, $\eta_{\rm c}$ for electrons has the similar meaning of $\eta_{\rm v}$ for holes.
It is clear that the electron mobility is increased for $(2\epsilon_{\|}+\epsilon_{\perp})=-0.08$, as oberserved by us in Fig.\,\ref{f11}.
\medskip

Our numerically calculated results for electron ($\mu_{\rm c}$) and hole ($\mu_{\rm c}$) mobilities are presented in Fig.\,\ref{f11}, along with their
comparisons with our experimental data.
In our model calculations, we have taken $T=300$\,K and the other model parameters can be found from Tables\ \ref{t1} and \ref{t2}.
The good agreement between our numerical calculated results and measured data strongly support the physical modeling present in this section.

\section{Summary and Conclusion }
\label{sec5}

The semiconductor processing, fabrication and the resulting carrier transport characteristics of MSM devices fabricated as wall like structures in silicon on insulator technology were reported. MSM device dark current,
DC photocurrents, and the time response of carrier transport were investigated. The resulting conducting channels were actually smaller than their physical dimensions, a result of depletion of carrier near the interfaces.
As the physical channel widths were reduced by oxidation, strain was produced near the interface and strained lattice became a significant portion of the conducting channel. The increase in mobilities for both holes and
electrons stemming from the strained silicon resulted in a dramatic increase in carrier mobility for both electrons and holes as the physical channel width was reduced from $200$\,nm to $20$\,nm. The theoretical model
incorporating the effects of strain present in these nanoscale MSM devices compared favorably with experimental results, showing that hole mobilities increased with decreasing $L$. Additionally, if these electron and hole mobilities can be retained with the application of gate electrodes, then this technique may yield a much simpler path towards high performance CMOS, both $n$-channel and $p$-channel, than current techniques for either
planer ultra-thin body FETs or FinFETs.

\begin{acknowledgements}
The authors would like to acknowledge the Air Force Research Laboratory, Space Vehicles Directorate for their support and interest in this work.
\end{acknowledgements}

\appendix

\section{Carrier Scattering Time}
\label{a1}

In general, the carrier concentration includes both the doping and photo-excitation contributions. If the sample is undoped, we can simply neglect the impurity scattering and have $n_{\rm c}=n_{\rm v}$.
The optical-phonon scattering and the inter-valley scattering are only important at high temperatures, while the acoustic-phonon scattering becomes more important at low temperatures.\,\cite{laux}
The surface-roughness scattering, on the other hand, is largely independent of temperature.
\medskip

For the impurity scattering, by using the Fermi's golden rule, its scattering rate $1/\tau_{\rm imp}$ is calculated as\,\cite{huang1,huang2}

\[
\frac{1}{\tau_{\rm imp}}=\frac{2}{N_{\rm c}}\sum_{{\bf k}}\,\frac{n_k}{\tau_{\rm imp}(k)}=\frac{2}{N_{\rm c}}\sum_{{\bf k}}\,n_k\left[N_{\rm i}\frac{2\pi}{\hbar}\sum_{{\bf q}}\,\left|\frac{-Ze^2}{\epsilon_0\epsilon_{\rm r}(q^2+Q_{\rm s}^2){\cal V}}\right|^2(1-n_{k+q})\,
\delta(E_{k+q}-E_k)\right]
\]
\begin{equation}
\approx\frac{n_{\rm i}Z^2e^4m^\ast}{2\pi\hbar^3\epsilon_0^2\epsilon^2_{\rm r}}\,\frac{2}{N_{\rm c}}\sum_{{\bf k}}\,n_k\,\frac{Q_{\rm s}^2}{k(4k^2+Q_{\rm s}^2)^2}\ ,
\end{equation}
where $N_{\rm c}$ is the total number of carriers in the system,
$n_{\rm i}=N_{\rm i}/{\cal V}$ is the impurity concentration, $Z$ is the impurity charge number, $\epsilon_{\rm r}=11.9$ is the silicon dielectric constant,
$Q_s^2=(e^2n_{\rm c}/\epsilon_0\epsilon_{\rm r}k_BT)$ at high temperatures with $n_{\rm c}=N_{\rm c}/{\cal V}$,
$E_k=\hbar^2k^2/2m^\ast$ is the carrier kinetic energy, and $m^\ast$ stands for the carrier effective mass.
For this case, we have $\alpha=1$.
In addition, at high temperatures we get conduction-band electron distribution

\begin{equation}
n^{\rm e}_k=\frac{1}{1+\exp[(E_k-u_{\rm c})/k_BT]}\approx\frac{n_{\rm c}}{2g_{\rm X}}\left(\frac{2\pi\hbar^2}{m_{\rm X}^\ast k_BT}\right)^{3/2}\exp\left(-\frac{E_k}{k_BT}\right)\ ,
\end{equation}
where we have assumed the high-energy $L$ valley becomes depopulated. Similar results can be obtained for valence-band hole distributions.
\medskip

For the longitudinal-acoustic-phonon scattering at high temperatures ($\hbar\omega_q\ll k_BT$), its scattering rate $1/\tau_{\rm ac}$ is calculated as\,\cite{huang1,huang2}

\[
\frac{1}{\tau_{\rm ac}}=\frac{2}{N_{\rm c}}\sum_{{\bf k}}\,\frac{n_k}{\tau_{\rm ac}(k)}=\frac{2}{N_{\rm c}}\sum_{{\bf k}}\,n_k\left\{\frac{2\pi}{\hbar}\sum_{{\bf q}}\,\frac{\hbar}{2\rho_0{\cal V}\omega_q}\left[D^2_{\rm ac}q^2+\frac{9}{32}(eh_{14})^2\right]
\left(\frac{q^2}{q^2+Q_s^2}\right)^2\right.
\]
\[
\times\left.\left[(1-n_{k+q})\,N_q\,\delta(E_{k+q}-E_k-\hbar\omega_q)
+(1-n_{k-q})\,(N_q+1)\,\delta(E_{k-q}-E_k+\hbar\omega_q)\right]\right\}
\]
\begin{equation}
\approx\frac{2\pi D^2_{\rm ac}k_BT}{\rho_0\hbar v_s^2}\,\frac{2}{N_{\rm c}}\sum_{{\bf k}}\,n_k\,g_{\rm 3D}(E_k)\ ,
\end{equation}
where $g_{\rm 3D}(E)=m^{\ast\,3/2}\sqrt{2E}/\pi^2\hbar^3$ is the three-dimensional density of states of carriers,
$N_q\equiv N_0(\hbar\omega_q/k_BT)$, $N_0(x)=1/[\exp(x)-1]$ is the Bose function for thermal-equilibrium phonons,
$\omega_q=v_sq$, $v_s=9\times 10^5$\,cm$/$s is the sound velocity, $\rho_0=2.33$\,g$/$cm$^3$ is the atomic mass density, $D_{\rm ac}=5.39$\,eV is the deformation potential for acoustic phonons, and $h_{14}$ is the piezoelectric constant neglected. For this case, we have $\alpha=3/2$.
\medskip

For the longitudinal-optical-phonon scattering, its scattering rate $1/\tau_{\rm op}$ is calculated as\,\cite{huang1,huang2}

\[
\frac{1}{\tau_{\rm op}}=\frac{2}{N_{\rm c}}\sum_{{\bf k}}\,\frac{n_k}{\tau_{\rm op}(k)}=\frac{2}{N_{\rm c}}\sum_{{\bf k}}\,n_k\left\{\frac{2\pi}{\hbar}\sum_{{\bf q}}\,\frac{\hbar\Omega_0}{2{\cal V}}\left(\frac{1}{\epsilon_{\infty}}-\frac{1}{\epsilon_{\rm s}}\right)
\,\frac{e^2}{\epsilon_0(q^2+Q_s^2)}\right.
\]
\[
\times\left.\left[(1-n_{k+q})\,N_{\rm LO}\,\delta(E_{k+q}-E_k-\hbar\Omega_0)
+(1-n_{k-q})\,(N_{\rm LO}+1)\,\delta(E_{k-q}-E_k+\hbar\Omega_0)\right]\right\}
\]
\begin{equation}
\approx\left(\frac{D_{\rm op}}{e\ell_{\rm op}}\right)^2\frac{e^2}{8\pi^2\rho_0\Omega_0}\,
\frac{2}{N_{\rm c}}\sum_{{\bf k}}\,n_k\left[(N_{\rm LO}+1)\,g_{\rm 3D}(E_k-\hbar\Omega_0)+N_{\rm LO}\,g_{\rm 3D}(E_k+\hbar\Omega_0)\right]
\end{equation}
where $N_{\rm LO}\equiv N_0(\hbar\Omega_0/k_BT)$, $\hbar\Omega_0=63$\,meV is the energy of optical phonons,
$(D_{\rm op}/e\ell_{\rm op})=2.2\times 10^{10}$\,V$/$m is the optical-polarization field.
For this case, we also have $\alpha=3/2$.
\medskip

For the surface-roughness scattering, its scattering rate $1/\tau_{\rm sr}$ is calculated as\,\cite{ando}

\begin{equation}
\frac{1}{\tau_{\rm sr}}=\frac{2}{N_{\rm c}}\sum_{{\bf k}}\,\frac{n_k}{\tau_{\rm sr}(k)}=\frac{m^\ast\Lambda^2e^4n_{\rm depl}}{\hbar^3\epsilon_0^2\epsilon_{\rm r}^2}\left(\frac{\delta b}{L}\right)^2
\frac{2}{N_{\rm c}}\sum_{{\bf k}}\,n_k\,\,\frac{1}{\sqrt{1+k^2\Lambda^2}}\,{\cal E}\left(\frac{k\Lambda}{\sqrt{1+k^2\Lambda^2}}\right)\ ,
\end{equation}
where $\delta b$ is the average roughness, $\Lambda$ is the roughness spatial-correlation length in a Gaussian model, and ${\cal E}(x)$ is the a complete elliptic integral. Additionally,
$(e/\epsilon_0\epsilon_{\rm r})\,n_{\rm depl}$ stands for the surface depletion-charge field, and $n_{\rm depl}$ is the surface depletion-charge areal densities.
For this case, we have $\alpha=1$.
\medskip

For the inter-valley scattering, its scattering rate $1/\tau_{\rm iv}$ can be calculated in a similar way for phonons, which gives

\[
\frac{1}{\tau_{\rm iv}}=\frac{2}{N_{\rm c}}\sum_{\xi}\,\frac{n^\xi_k}{\tau^\xi_{\rm iv}(k)}=
\sum_{\xi,\,\xi^\prime}\left(\frac{D_{\xi\xi^\prime}}{e\ell_{\xi\xi^\prime}}\right)^2\frac{e^2}{8\pi^2\rho_0\omega_{\xi\xi^\prime}}\,
\frac{2}{N_{\rm c}}\sum_{{\bf k}}\,n^\xi_k
\]
\begin{equation}
\times\left\{[N(\omega_{\xi\xi^\prime})+1]\,g_{\rm 3D}^\xi(E^\xi_k-\Delta E_{\xi\xi^\prime}-\hbar\omega_{\xi\xi^\prime})
+N(\omega_{\xi\xi^\prime})\,g_{\rm 3D}^\xi(E^\xi_k-\Delta E_{\xi\xi^\prime}+\hbar\omega_{\xi\xi^\prime})\right\}\ ,
\end{equation}
where $(D_{\xi\xi^\prime}/e\ell_{\xi\xi^\prime})$ is the inter-valley optical-polarization field, $N(\omega_{\xi\xi^\prime})\equiv N_0(\hbar\omega_{\xi\xi^\prime}/k_BT)$,
$\omega_{\xi\xi^\prime}=v_s|{\bf K}_{\xi^\prime}-{\bf K}_{\xi}|$, and $\Delta E_{\xi\xi^\prime}=E^{\xi^\prime}_{\rm G}-E^{\xi}_{\rm G}$.
For this case, we have $\alpha=3/2$.
\medskip

The finite-size effect in the direction perpendicular to the silicon film becomes significant as $\pi^2\hbar^2/2m^\ast_{\rm X}L^2\gg k_BT$.\,\cite{huang3}
The existence of such a quantum well modify the splitting of heavy and light holes by $E_{\rm HH}\rightarrow E_{\rm HH}+\Delta^{\rm v}_{\rm qw}$ and
$E_{\rm HH}\rightarrow E_{\rm LH}-\Delta^{\rm v}_{\rm qw}$, where $2\Delta^{\rm v}_{\rm qw}$ stands for the quantum-well induced valence-band splitting, as well as $g_{\Gamma}\rightarrow 1$.
It also affects the bandgap energy by $\varepsilon_{\rm G}^{\rm X}(T)\rightarrow\varepsilon_{\rm G}^{\rm X}(T)+\Delta^{\rm c}_{\rm qw}$, as well as the density of states of carriers
by $g_{\rm 3D}(E_k)\propto\sqrt{E_k}\rightarrow g_{\rm 2D}(E_k)\propto{\rm constant}$. Additionally, the coulomb potential in the momentum space is changed by
$e^2/\epsilon_0(q^2+Q_s^2){\cal V}\rightarrow e^2/\epsilon_0(q+q_s){\cal A}$, where ${\cal A}$ is the area of the quantum well and $1/q_s$ is the Thomas-Fermi screening length for quantum wells.
It is clear that the film quantization effect tends to reduce the strain-induced mobility enhancements of both electrons and holes.

\newpage
\begin{table}
\caption{Model parameters used in calculating mobility of electrons in strained Si film.}
\begin{tabular}{cccc}
  \hline
    $\mu_{\rm c}^{\rm max}$ ($cm^2/V\cdot s$)\ \ \ \ &
    $\mu_{\rm c}^{(0)}$ ($cm^2/V\cdot s$)\ \ \ \ &
    $\lambda_{\rm c}$ (nm)\ \ \ \ &
    $2{\cal D}_{\rm c}$ (nm)\\
  \hline\hline
    $5500$ & $806$ & $15$ & $42$\\
  \hline
\end{tabular}
\label{t1}
\end{table}

\begin{table}
\caption{Model parameters used in calculating mobility of holes in strained Si film.}
\begin{tabular}{cccc}
  \hline
    $\mu_{\rm v}^{\rm max}$ ($cm^2/V\cdot s$)\ \ \ \ &
    $\mu_{\rm v}^{(0)}$ ($cm^2/V\cdot s$)\ \ \ \ &
    $\lambda_{\rm v}$ (nm)\ \ \ \ &
    $2{\cal D}_{\rm v}$ (nm)\\
  \hline\hline
    $3000$ & $100$ & $143$ & $42$\\
  \hline
\end{tabular}
\label{t2}
\end{table}

\end{document}